\documentclass[lettersize,journal]{IEEEtran}
\usepackage{amsmath,amsfonts}
\usepackage{algorithmic}
\usepackage{algorithm}
\usepackage{array}
\usepackage[caption=false,font=normalsize,labelfont=sf,textfont=sf]{subfig}
\usepackage{textcomp}
\usepackage{stfloats}
\usepackage{url}
\usepackage{verbatim}
\usepackage{graphicx}
\usepackage{cite}
\usepackage{multirow}
\usepackage{adjustbox}
\usepackage{makecell}
\usepackage{mathtools}
\usepackage{comment}
\usepackage[normalem]{ulem}
\usepackage[inline]{enumitem}
\usepackage{soul}
\usepackage{bm}
\usepackage{siunitx}
\usepackage{xcolor}
\usepackage{subfig}
\usepackage{amssymb}
\usepackage{tablefootnote}

\usepackage[font=small]{caption}
\setstcolor{red}

\newcolumntype{C}[1]{>{\centering\arraybackslash}m{#1}}

\begin{document}

\title{Quantum-Assisted Joint Virtual Network Function Deployment and Maximum Flow Routing for Space Information Networks}

\author{Yu~Zhang,~\IEEEmembership{Student~Member,~IEEE,}
        Yanmin~Gong,~\IEEEmembership{Senior~Member,~IEEE,}
        Lei~Fan,~\IEEEmembership{Senior~Member,~IEEE,}
        Yu~Wang,~\IEEEmembership{Fellow,~IEEE,}
        Zhu~Han,~\IEEEmembership{Fellow,~IEEE}, and Yuanxiong~Guo,~\IEEEmembership{Senior~Member,~IEEE}
\thanks{Y. Zhang and Y. Gong are with the Department of Electrical and Computer Engineering, University of Texas at San Antonio, Texas 78249, USA. (e-mail: \{yu.zhang@my., yanmin.gong@\}utsa.edu).}
\thanks{L. Fan is with the Department of Engineering Technology and Department of Electrical and Computer Engineering, University of Houston, Houston, Texas 77204, USA. (e-mail: lfan8@central.uh.edu).}
\thanks{Y. Wang is with the Department of Computer and Information Sciences, Temple University, Philadelphia, Pennsylvania 19122, USA. (e-mail: wangyu@temple.edu).}
\thanks{Z. Han is with the Department of Electrical and Computer Engineering, University of Houston, Houston, Texas 77204, USA, and also with the Department of Computer Science and Engineering, Kyung Hee University, Seoul 446-701, South Korea. (e-mail: zhan2@uh.edu).}
\thanks{Y. Guo is with the Department of Information Systems and Cyber Security, University of Texas at San Antonio, Texas 78249, USA. (e-mail: yuanxiong.guo@utsa.edu).}
\thanks{The work is partially supported by the US NSF (Grant NO.  ECCS-2045978, CNS-2006604, CNS-2128378, OAC-2417716, CNS-2107216, CNS-2128368, CMMI-2222810, ECCS-2302469.), US Department of Transportation, Toyota, Amazon, and Japan Science and Technology Agency (JST) Adopting Sustainable Partnerships for Innovative Research Ecosystem (ASPIRE) under JPMJAP2326. (Corresponding author: Y. Guo.)}
}



\maketitle

\begin{abstract}
Network function virtualization (NFV)-enabled space information network (SIN) has emerged as a promising method to facilitate global coverage and seamless service. This paper proposes a novel NFV-enabled SIN to provide end-to-end communication and computation services for ground users. Based on the multi-functional time expanded graph (MF-TEG), we jointly optimize the user association, virtual network function (VNF) deployment, and flow routing strategy (U-VNF-R) to maximize the total processed data received by users. The original problem is a mixed-integer linear program (MILP) that is intractable for classical computers. Inspired by quantum computing techniques, we propose a hybrid quantum-classical Benders' decomposition (HQCBD) algorithm. Specifically, we convert the master problem of the Benders' decomposition into the quadratic unconstrained binary optimization (QUBO) model and solve it with quantum computers. To further accelerate the optimization, we also design a multi-cut strategy based on the quantum advantages in parallel computing. Numerical results demonstrate the effectiveness and efficiency of the proposed algorithm and U-VNF-R scheme.  
\end{abstract}


\begin{IEEEkeywords}
Space information network, multi-functional time expanded graph, quantum computing, Benders' decomposition.
\end{IEEEkeywords}

\section{Introduction} 
\label{sec: introduction}  

\IEEEPARstart{I}{n} the era of sixth-generation (6G) communication, space information network (SIN) is an important component for providing seamless global coverage. SINs enable real-time data acquisition \cite{zhang2008satellite}, extensive data transmission and processing \cite{zhang2024energy, yu2020joint}, and systematized information services \cite{al2021multi}. Leveraging these advantages, an increasing number of innovative applications are emerging, such as ocean-going voyages and hologram video streaming. However, conventional SINs are usually tailored to support particular tasks (e.g., Earth observation or communication). It is challenging for conventional SINs to manage the burgeoning variety of new applications.  

To meet the diverse service requirements of emerging applications, network function virtualization (NFV) is a promising approach. In NFV, network functions (e.g., firewalls and load balancers), which traditionally require dedicated physical hardware, are transformed into software components known as virtual network functions (VNFs) \cite{mijumbi2015network}. Then, VNFs can be flexibly deployed on different physical nodes to establish a customized virtual network for each requested service \cite{ye2016joint}. In this way, NFV can provide significant flexibility and enhance the efficiency of resource utilization. Utilizing the NFV technology, the SIN can support various applications concurrently. In the NFV-enabled SIN, each task is characterized by a specific and predefined service function chain (SFC). For the effective processing of the task, it is crucial to ensure that the task flow satisfies the SFC constraints. In other words, the VNFs associated with the task are processed sequentially from the source node to the destination node, following the order that has been predetermined \cite{yang2021maximum}. 

There are two inherent challenges in NFV-enabled SINs. The first is the strategic deployment of VNFs to optimize service efficiency. The second involves developing a flow routing strategy that aligns with SFC constraints. There is a substantial amount of literature to tackle these challenges. However, most existing works either focus on flow routing strategies in NFV-enabled SINs while neglecting VNFs deployment \cite{wang2018multi, hu2020joint, jia2021toward} or just consider each satellite configured with a single VNF deployment \cite{gao2022virtual, jia2021vnf}. Recently, a few studies have tried to jointly study the flow routing strategy and VNFs deployment, considering configurations where each satellite supports the deployment of multiple VNFs \cite{yang2023group, yang2023space}. However, in these studies, the researchers only consider Earth observation scenarios and study the flow routing strategies among satellites. Their solution cannot be directly applied to multi-user NFV-enabled SINs since the connectivity between satellites and ground users is often limited by their respective communication capacities. With increasing interest from companies like SpaceX in leveraging satellites for end-to-end user services \cite{spaceX_apple}, there is a pressing need to extend these models. Moreover, unlike the static network topology of terrestrial networks, the network topology of SINs not only is time-varying but also comprises thousands of nodes. This leads to a heightened requirement for collaboration and coordination among the various components. Therefore, the joint user association, VNF deployment, and maximum flow routing strategy (U-VNF-R) problem has to be formulated as a large-scale mixed-integer linear program (MILP) in practice, which is NP-hard. Consequently, it is difficult to utilize classical computing techniques to solve such problems.

Quantum computing (QC) offers a promising approach to solving large-scale combinatorial optimization problems. Unlike classical computing, which relies on binary bits, QC utilizes qubits to encode superposition states and leverage quantum entanglement and tunneling, enabling the simultaneous exploration of exponential state combinations \cite{bharti2022noisy}. One notable QC technique is quantum annealing (QA), typically implemented in specialized machines known as quantum annealers. QA offers a larger number of qubits compared to other QC methods \cite{gyongyosi2019survey}. With these qubits, QA can address real-world problems such as satellite caching resource allocation \cite{zhang2023quantum} and beam placement \cite{dinh2023efficient}. However, the major limitation of QA is that it only accepts quadratic unconstrained binary optimization (QUBO) formulation. Owing to this reason, most prior studies typically formulate real-world applications as binary quadratic model (BQM) \cite{QA_RNA1,QA_RNA2,QA_car} or integer linear programming (ILP) \cite{dinh2023efficient, doan2022hybrid} that can be readily transformed into the QUBO formulation. However, BQM and ILP formulations often fail to capture the complexity of problems in the NFV-enabled SIN field \cite{al2022survey, wang2019convergence, fang20215g}.

To overcome the aforementioned challenges, this paper introduces an innovative hybrid quantum-classical approach to efficiently solve the complex joint optimization problem in a novel NFV-enabled SIN. In this system, the NFV-enabled SIN serves a dual role, providing both end-to-end communication and computational services to ground users. Specifically, the ground users in the source city offload their individual tasks to the NFV-enabled SIN through the associated satellite. Then, the NFV-enabled SIN processes them in a sequential manner, according to their SFCs. Upon completion, the processed results are then transmitted to the target users in the destination city. However, it is challenging to represent the resources and demands for the NFV-enabled SIN. According to recent research in SINs \cite{liu2023multi}, we adopt the multi-functional time expanded graph (MF-TEG) to depict the communication, storage, and computation capability of the dynamic SIN over time. Compared with traditional snapshot sequence graph (SSG) and time expanded graph (TEG), MF-TEG not only skills in modeling the spatial and temporal connections for dynamic networks but also can uniquely characterize the capability where multiple VNFs of one task flow can be simultaneously deployed on a single node. Based on MF-TEG, we focus on maximizing the total processed data received by the ground user in the destination city by jointly optimizing the user association, VNFs deployment, and flow routing strategy under multiple resource constraints and flow restrictions. 

The formulated optimization problem is a large-scale MILP which poses a significant challenge for classical computers to solve. Additionally, it is not feasible to discretize all continuous variables and directly implement QC due to the constraints imposed by the limited number of available qubits. Therefore, we propose a hybrid quantum-classical Benders' decomposition (HQCBD) algorithm. In particular, based on the Benders' decomposition (BD), the original MILP is decomposed into a master problem with the binary decision variables and a subproblem with the continuous decision variables. On one hand, the subproblem is linear programming (LP), which can be directly solved by LP solvers on classical computers. On the other hand, the master problem is a large-scale ILP problem, which is NP-hard. Therefore, we utilize the constraint-penalty pair principle \cite{zhao2022hybrid} to reformulate the master problem into the QUBO formulation that is solvable by QA. Then, we iteratively solve these two problems until their solutions converge. Additionally, we recognize that the parallel computing capabilities of quantum computers are extremely powerful. Therefore, we have designed a specialized quantum multi-cut strategy to accelerate HQCBD. Finally, the proposed HQCBD algorithm and U-VNF-R scheme are evaluated on the D-Wave's real-world quantum computer \cite{dwave}.

To recapitulate, the main contributions of this paper are summarized as follows:
\begin{itemize}
    \item We propose a cutting-edge NFV-enabled SIN where satellites cooperatively provide end-to-end communication and computation services for ground users. To clearly depict the dynamic topology of NFV-enabled SIN, we leverage the MF-TEG to represent the resources and demands in the SIN, which facilitates the formulation and algorithm design. 
    \item We introduce a novel hybrid quantum-classical algorithm named HQCBD to effectively solve the formulated NFV-enabled SIN model, which is characterized as a large-scale MILP. Furthermore, we have enhanced the proposed algorithm with a multi-cut strategy to further accelerate its convergence.
    \item We conduct extensive simulations using the D-Wave quantum computer. Numerous results have demonstrated the superiority of our proposed HQCBD compared with the classical computing algorithm. Moreover, these results also reveal that our proposed U-VNF-R scheme significantly outperforms baselines in various aspects.
\end{itemize}

The rest of this paper is organized as follows. Section~\ref{sec: related works} discusses the related works. Section \ref{QA} introduces the preliminaries of QA. Section \ref{sec: model} describes the system model and problem formulation. In Section \ref{sec: solution}, we present the HQCBD algorithm to solve the formulated problem. Section~\ref{sec: experiment} shows the simulation results. Finally, Section \ref{sec: conclusion} concludes the paper.

\section{Related Works} 
\label{sec: related works}

\subsection{NFV-enabled SIN}
NFV-enabled SINs have attracted considerable research interest in recent years due to their potential benefits in advancing 6G and beyond networks \cite{jiang2021road}. Substantial literature exists to investigate the flow routing strategy in NFV-enabled SINs \cite{wang2018multi, hu2020joint, jia2021toward}. Based on the TEG, Wang et al. \cite{wang2018multi} investigated a flow routing strategy aiming to maximize the sum of priorities for successfully completed tasks while satisfying the latency constraints in the SIN. Based on the SSG, Hu et al. \cite{hu2020joint} optimized the flow routing strategy to maximize data rates while meeting the signal-to-interference-plus-noise ratio requirements in the SIN, given limited available time resources. Based on the TEG, Jia et al. \cite{jia2021toward} studied the flow routing strategy to maximize the total data received by the ground data processing center under multiple resource constraints and flow restrictions. However, the above works mainly focus on optimizing flow routing strategies in NFV-enabled SINs while neglecting the VNFs deployment. There is also a variety of literature investigating both the deployment of VNFs and flow routing strategies in SINs \cite{gao2022virtual, jia2021vnf}. Based on the SSG, Gao et al. \cite{gao2022virtual} studied the joint VNFs deployment and flow routing strategy to maximize the overall network payoff while meeting the end-to-end latency requirements in the SIN. Based on the TEG, Jia et al. \cite{jia2021vnf} jointly optimized the VNF deployment and flow routing strategy to minimize resource consumption. Nevertheless, these studies employ traditional TEG and SSG approaches, which restrict their scope to configurations where each satellite is equipped with only a single VNF deployment. Only a few recent studies tried to jointly optimize the VNF deployment and flow routing strategy, considering configurations where each satellite supports the deployment of multiple VNFs \cite{yang2023group, yang2023space}. Based on the MF-TEG, Yang et al. \cite{yang2023group} studied the trade-off between the network maximum flow and coordination overhead under the SFC constraints. Based on the MF-TEG, Yang et al. \cite{yang2023space} jointly optimized the VNF deployment and flow routing strategy, aiming at maximizing the overall network payoff while satisfying the latency requirements in the SIN. 

However, these studies focus on flow routing strategies among satellites, which are not directly applicable to multi-user NFV-enabled SINs. This limitation arises from the fact that the communication capacities of SINs are often constrained by the connectivity between satellites and ground users. Besides, most of the existing works in the NFV-enabled SIN systems usually formulate their problems as large-scale MILPs, which are hard to solve. Therefore, they either use heuristics or complex optimization techniques to tackle them. In view of the above limitations, we propose an innovative NFV-enabled SIN to provide end-to-end communication and computation services for ground users. Then, we design an HQCBD algorithm to solve the formulated MILP efficiently.

\subsection{QA for Optimization}
QC encompasses two main paradigms: gate-based QC and adiabatic QC (AQC). Gate-based QC uses discrete quantum gate operations but is limited by circuit depth and qubit count, with current implementations providing fewer than 150 qubits \cite{IBM}. AQC, on the other hand, encodes problems into the Hamiltonian of a quantum system, seeking optimal solutions in its ground states, but is challenging due to system vulnerabilities. QA is considered a form of relaxed AQC, which does not strictly require universality or adiabaticity \cite{kadowaki1998quantum}. Nowadays, more than 5,000 qubits are available for QA\cite{dwave}.

Recently, extensive research efforts have been made to utilize QA for solving practical optimization problems \cite{fan2022hybrid}. However, a significant limitation of QA lies in its exclusive acceptance of the QUBO formulation. Thus,  most prior works typically formulate real-world applications as BQM  \cite{QA_car,QA_RNA1,QA_RNA2} or ILP problems \cite{doan2022hybrid,dinh2023efficient}, which can be easily transformed into QUBO formulation. For instance, Fox et al. \cite{QA_RNA1} implemented the codon optimization utilizing a BQM. Mulligan et al. \cite{QA_RNA2} formulated the protein design problem as a BQM, aiming to find amino acid side chain identities and conformations to stabilize a fixed protein backbone. \cite{QA_car} formulated a BQM problem to minimize the number of color switches between cars in a paint shop queue. Doan et al. \cite{doan2022hybrid} formulated the robust fitting as an ILP problem and found the global or tightly error-bounded solution utilizing QA. Dinh et al. \cite{dinh2023efficient} modeled satellite communication beam placement as an ILP and introduced an efficient Hamiltonian Reduction approach for QA to solve this problem.  

However, since SINs generally involve many continuous resources, the BQM and ILP formulations are insufficient to fully encompass the complexity inherent in the NFV-enabled SIN field \cite{al2022survey, wang2019convergence, fang20215g}. To the best of our knowledge, this is the first work to utilize QA for jointly optimizing the user association, VNFs deployment, and flow routing strategy in the NFV-enabled SIN system, which is formulated as a MILP.


\section{PRELIMINARIES}
\label{QA}
QA surpasses classical computing in solving large-scale combinatorial optimization problems \cite{QA_car,QA_RNA1,QA_RNA2, doan2022hybrid,dinh2023efficient}. The combinatorial optimization problems typically focus on minimizing a cost function, which can be equivalently converted into finding the ground state of a classical Ising Hamiltonian, represented as $H_p$ \cite{lucas2014ising}. Nevertheless, many formulated optimization problems have multiple local minima, corresponding to Ising Hamiltonians that are reminiscent of classical spin glasses. These characteristics greatly hinder classical algorithms in finding the global minimum \cite{kadowaki2002study}. QA emerged as a promising alternative to tackle this formidable task. In QA, the classical Ising Hamiltonian $H_p$ is transformed into the quantum domain and described by a collection of interacting qubits. We briefly introduce the theoretical background of QA as follows. 

Based on the adiabatic theorem of quantum mechanics \cite{amin2009consistency},  the quantum system’s state can be initialized in the ground state of the initial Hamiltonian $H_0$. Subsequently, we gradually evolve the system's state towards the target Hamiltonian $H_P$. This process is carefully managed to guarantee that the system remains in the ground state consistently during the entire evaluation period. Finally, by measuring the final ground state of the system, we can obtain the ground state of the target Hamiltonian $H_p$. This state also represents the solution to the original optimization problem\cite{yarkoni2022quantum}. Let the evolution time be denoted by $\tau_e \in [0, T_e]$, and let $K$ represent the number of qubits. The time-dependent evolution can be described as 
\begin{equation} \label{eq: quantum}
    H(\tau_e) = A(\tau_e)H_0 + B(\tau_e)H_p,
\end{equation}
where
\begin{align}
& H_0=\sum_{i = 1}^K\sigma^x_i, \quad H_p=\sum_{i  = 1}^Kh_i\sigma^z_i + \sum_{i,j =1}^KJ_{i,j}\sigma^z_i\sigma^z_j.
\end{align}
Here, $\sigma_i^x$ is the $x$-Pauli matrix, while $\sigma_i^z$ is the $z$-Pauli matrix, acting on the $i$-th qubit \cite{lucas2014ising}. $J_{i,j} = J_{j,i}$ is called coupling strength, which represents the symmetric interaction strength of the qubits $i$ and $j$.  $h_i$ is called bias, which describes the on-site energy of qubit $i$. In the evolution, $A(\tau_e)$ is slowly reduced from the initial value $A(0) = 1$ to the final value $A(T_e)=0$ and $B(\tau_e)$ is slowly increased from the initial value $B(0) = 0$ to the final value $B(T_e)=1$. Therefore, the Hamiltonian is changing from $H(0)=H_0$ to $H(T_e)=H_p$,  from which the global optimal solution can be derived. 

According to \cite{das2005quantum},  the quantum $\sigma^z$ Pauli operators can be equivalently substituted by the classical spin variables in Hamiltonian $H_p$. Then, we can obtain an Ising model representing the Hamiltonian $H_p$, i.e., 
\begin{align}  \label{eq: Ising}
& \min_{\textbf{s} \in \{-1,1\}^K} H_p(\textbf{s})=\sum_{i=1}^Kh_is_i + \sum_{i=1}^{K}\sum_{j=1}^{K}J_{i,j}s_is_j.
\end{align}

Alternatively, we can express the Ising model as a QUBO formulation. Let $f_Q:\{0,1\}^K \rightarrow \mathbb{R}$ be a quadratic polynomial over binary variables $\mathbf{x}=[x_1,\dots,x_K]$, and $\mathbf{Q}\in \mathbb{R}^{K\times K}$ be an upper triangular matrix. The QUBO formulation is given as \cite{scherer2019mathematics}
\begin{equation} \label{QUBO_1}
\min_{\textbf{x}\in \{0,1\}^K}f_Q(\mathbf{x})=\sum_{i=1}^KQ_{ii}x_i + \sum_{i=1}^{K}\sum_{j=1}^{K}Q_{ij}x_ix_j=\mathbf{x}^\intercal\mathbf{Q}\mathbf{x}.
\end{equation}
Note that the QUBO formulation can be easily transformed back into the Ising model by mapping $x_i = \frac{s_i+1}{2}$.

\section{System Model and Problem Formulation} 
\label{sec: model}
In this section, we first introduce the NFV-enabled SIN. Next, we present the MF-TEG model for the time-varying NFV-enabled SIN. Finally, we formulate an optimization problem to maximize the total processed data received by users.

\subsection{NFV-enabled SIN}
\label{subsec:system_model}
\begin{figure}[t]
\centering
\includegraphics[scale=0.26]{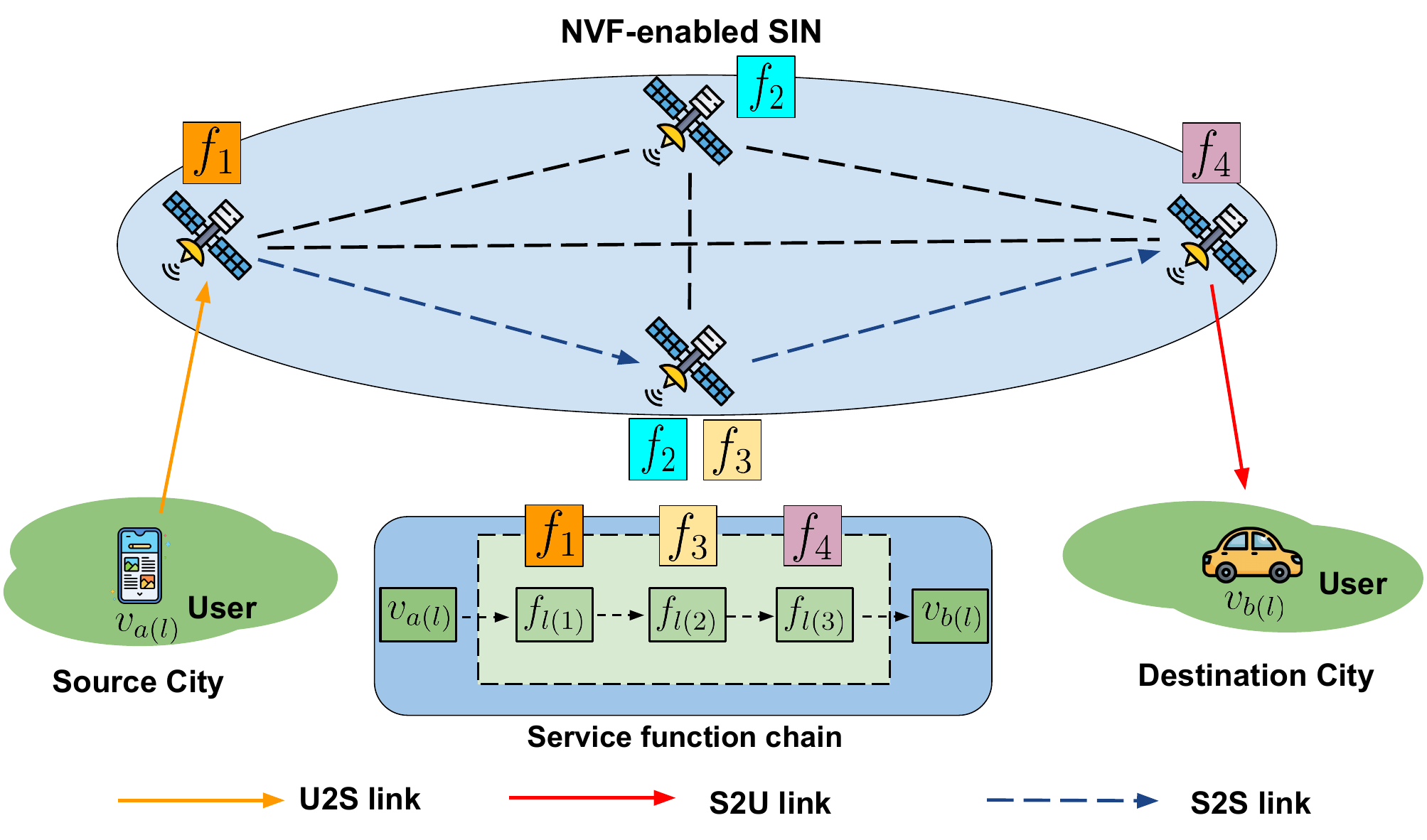}
\caption{An example of the SFC with three functions in the SIN.}
\label{fig:system_model}
\end{figure}

As shown in Fig. \ref{fig:system_model}, we consider an NFV-enabled SIN, which comprises a set of satellites $\mathcal{S}$. Each satellite is equipped with an edge server and connects to other satellites via Satellite-to-Satellite (S2S) links. These links are maintained through adaptive link management techniques like smart antennas and beamforming \cite{radhakrishnan2016survey}. In this NFV-enabled SIN, we suppose there are $L$ task flows from the source city to the destination city, which is denoted as a set $l \in \mathcal{L}=\{1, \ldots, L\}$. Each task flow $l$ can be characterized by a tuple $l = \{v_{a(l)}, v_{b(l)}\}$, where $v_{a(l)}, v_{b(l)}$ represent the source user, destination user of task flow $l$, respectively. Specifically, the source user $a(l)$ first transmits its task to the associated satellite via the User-to-Satellite (U2S) communication link. Then, the task is processed according to the service requirements during the routing process. Upon completion, the result is transmitted to the corresponding user in the destination city through an associated satellite via the Satellite-to-User (S2U) communication link. The requested services for each task flow are characterized by a service function chain (SFC), which comprises a set of functions performed in a predefined order. We denote the total required functions for all task flows as $ \mathcal{F} = \{f_c | 1 \leq c \leq F\}$. Furthermore, the SFC of task flow $l$ is denoted as $\mathcal{F}(l) =\{ f_{l(1)} \rightarrow \dots \rightarrow f_{l(k)} \rightarrow \dots \rightarrow f_{l(K_l)}\}$, where $K_l$ and $f_{l(k)} \in \mathcal{F}$ represents the total number of functions and the $k$-th required function of task flow $l$, respectively. We assume there is a central network controller equipped with both classical and quantum computers to manage user association, VNFs deployment, and flow routing strategy for the SIN. Table. \ref{table: notation} is the summary of key notations in this work.

\begin{table}[t!] 

\caption{List of Notations}
\label{table: notation}
\centering
\begin{adjustbox}{width=\columnwidth,center}
\begin{tabular}{|C{2.1cm}|C{6.5cm}|}
\hline
\textbf{Notation} & \textbf{Definitions}  \\
\hline
$\mathcal{L}$; $L$; $l$ & the set of task flows; the number of the task flows; the $l$-th task flow\\
\hline
$v_{a(l)}; v_{b(l)}$ & the source user of task flow $l$; the destination user of task flow $l$ \\
\hline
$\mathcal{F}; f_c; F$ & the set of functions for all task flows; the $c$-th function; the number of the functions in $\mathcal{F}$\\
\hline
$\mathcal{F}(l); f_{l(k)}; K_l$ & the SFC of task flow $l$; the $k$-th function required by task flow $l$; the number of functions in  $\mathcal{F}(l)$\\
\hline
\makecell{$\mathcal{V};\mathcal{A};\mathcal{V}_M;$\\$\mathcal{A}_M$} & the set of nodes; the set of links; the set of nodes in the MF-TEG; the set of links in the MF-TEG \\
\hline
$\mathcal{V}_a;\mathcal{V}_b;\mathcal{V}_s$ & the set of source users; the set of destination users; the set of satellites \\
\hline
$\mathcal{V}_c;\mathcal{V}_{nc}$ & the set of function nodes; the set of non-function nodes\\
\hline
$\mathcal{F}_i; f_{i,n}; N_i$ & the set of functions decomposed by function node $v_i \in \mathcal{V}_c$; the $n$-th function of function node $v_i \in \mathcal{V}_c$; the total function number of function node $v_i \in \mathcal{V}_c$ \\
\hline
\makecell{$\mathcal{V}_{vs};\mathcal{V}_{vc};$\\$ \mathcal{V}_{vc, i}$} & the set of virtual sub-nodes; the set of virtual function nodes; the set of virtual functions decomposed
by the function node $v_i \in \mathcal{V}_c$ \\
\hline
$v_{i_0}; v_{i_{f_{i,n}}}$ & the virtual sub-node and $n$-th virtual function nodes decomposed by function node $v_i \in \mathcal{V}_c$, respectively  \\
\hline
$\mathcal{T}; T ; \tau$ & the set of time slots; the time horizon \\
\hline
\makecell{$v_{a(l)}^t; v_{b(l)}^t;$\\$ v_{i_0}^t; v_{i_{f_{i,n}}}^t$ }& the replicas of $v_{a(l)}, v_{b(l)}, v_{i_0}, v_{i_{f_{i,n}}}$ within the time $t$ in the MF-TEG, respectively \\
\hline
\makecell{$\mathcal{V}_{M, a};\mathcal{V}_{M, b};$\\$\mathcal{V}_{M, nc};\mathcal{V}_{M, vs};$\\$\mathcal{V}_{M, vc}$ }& the sets of replicas of source users, destination users, non-function nodes, virtual sub-nodes, and virtual function nodes in the MF-TEG, respectively \\
\hline
\makecell{$\mathcal{A}_{M, tr};\mathcal{A}_{M,vl};$\\$\mathcal{A}_{M,st}$ }& the sets of replicas of transmission links, virtual transmission links, and storage links in the MF-TEG, respectively  \\
\hline
\makecell{$\mathcal{A}_{M, vl}^{\text{in}}; \mathcal{A}_{M, vl}^{\text{out}}$ }& the set of virtual transmission links from virtual sub-node to the virtual function node in MF-TEG; the set of virtual transmission links from the virtual function node to virtual sub-node  \\
\hline
$\xi_{f_{l(k)}}$ & a virtual flow indicating that task flow $l$ just receiving function $f_{l(k)}$  \\
\hline
$C(v_i^t, v_j^t)$ & the communication capacity of the transmission link $(v_i^t, v_j^t)$ \\
\hline
\makecell{$C(v_{i_0}^t, v_{i_{f_{i,n}}}^t);$\\ $C(v_{i_{f_{i,n}}}^t, v_{i_0}^t)$} & the communication capacity of the virtual transmission links $(v_{i_0}^t, v_{i_{f_{i,n}}}^t)$ and $(v_{i_{f_{i,n}}}^t, v_{i_0}^t)$, respectively  \\
\hline
$C(v_i^t, v_i^{t+1})$ & the storage capacity of the storage link $(v_i^t, v_i^{t+1})$  \\
\hline
$K_{i, j}^t$  & the binary connectivity indicator indicating whether there is a connection between nodes $v_i^t$ and $v_j^t$ within time $t$\\
\hline
$H_{i_{f_{i,n}}, f^\prime}$ & the binary indicator indicating whether the virtual function node $v_{i_{f_{i,n}}} \in \mathcal{V}_{vc}$ provides function $f^\prime \in \mathcal{F}$ \\
\hline
$\beta_{\xi_{f_{l(k-1)}}, \xi_{f_{l(k)}}}$ & the scaling factor representing the ratio between the data amounts of $\xi_{f_{l(k-1)}}$ and $\xi_{f_{l(k)}}$  \\
\hline
$C_{i}^{\text{U2S}}; C_{i}^{\text{S2U}}$ & the maximum number of source users and destination users that satellite $i$ can connect, respectively  \\
\hline
$C_i^t$ & the computation capacity of function node $v_i \in \mathcal{V}_c$  \\
\hline
\makecell{$x_{i,j}^t(\xi_{f_{l(k)}})$;\\$ y_{i_0,i_{f_{i,n}}}^t(\xi_{f_{l(k)}})$;\\$ z_{i_{f_{i,n}}, i_0}^t(\xi_{f_{l(k)}})$} & the cumulative amount of data transmitted on transmission link $(v_i^t, v_j^t) \in \mathcal{A}_{M, tr}$, virtual input transmission link $(v_{i_0}^t, v_{i_{f_{i,n}}}^t) \in \mathcal{A}_{M, vl}^{\text{in}}$, and virtual output transmission link $(v_{i_{f_{i,n}}}^t, v_{i_0}^t) \in \mathcal{A}_{M, vl}^{\text{out}}$ for the virtual flow $\xi_{f_{l(k)}}$ within time $t$, respectively  \\
\hline
$o_i^{t, t+1}$ & the total amount of data stored on storage link $(v_i^t, v_i^{t+1}) \in \mathcal{A}_{M, st}$ from the time $t$ to $t+1$ for the virtual flow $\xi_{f_{l(k)}}$  \\
\hline
$\lambda_{i_{f_{i,n}}, f_{l(k)}}$ & the binary variables indicating whether the virtual function node $v_{i_{f_{i,n}}}$ serves function $f_{l(k)}$ for the task flow $l$  \\
\hline
$\phi^{\text{U2S},t}_{a(l), i};\phi^{\text{S2U},t}_{i, b(l)}$ & the binary variables indicating the associations between source user $v_{a(l)}$, destination user $v_{b(l)}$ and satellite node $ v_i \in \mathcal{V}_s$ for task flow $l$ at time $t$, respectively \\
\hline
\end{tabular}
\end{adjustbox}
\end{table}

In this paper, the task flows represent the streaming data (e.g., 4k live) sent by the source users to destination users. We are interested in maximizing the network throughput while ensuring each task flow from the source user to the destination user is processed by the required functions (e.g., encode, compress, decode) in the predefined order. This system can be described by a graph $\mathcal{G}= (\mathcal{V}, \mathcal{A})$. Here, vertices $\mathcal{V}=\mathcal{V}_a\cup\mathcal{V}_b\cup \mathcal{V}_s$, where $\mathcal{V}_a= \{v_{a(l)}| l \in \mathcal{L}\}$, $\mathcal{V}_b=\{v_{b(l)}| l \in \mathcal{L}\}$, and $\mathcal{V}_s = \{v_{i} | i \in \mathcal{S}\}$ represent the replicas of the set of source user nodes, destination user nodes, and satellite nodes in the NFV-enabled SIN, respectively. The links $\mathcal{A} = \{(v_i,v_j) |v_i, v_j \in \mathcal{V}, v_i \neq v_j \}$ indicate the transmission link from nodes $v_i$ to $v_j$.

Based on whether a satellite node has a function, satellite nodes can be classified into two categories: function nodes and non-function nodes. Function nodes are represented by the set $ \mathcal{V}_c = \{ v_i | v_i \text{ provides one or more functions for } L \text{ task } \allowbreak \text{flows} \} $, while non-function nodes are denoted by $\mathcal{V}_{nc} = \{ v_i | \allowbreak v_i  \text{ cannot } \text{provide any functions for } L \text{ task flows} \}$. Then, the entire set of nodes $\mathcal{V}$ can also be expressed as a union set of source user set $\mathcal{V}_a$, destination user set $\mathcal{V}_b$, the function satellite set $\mathcal{V}_c$, and the non-function satellite set $\mathcal{V}_{nc}$. It is written as $\mathcal{V} = \mathcal{V}_a \cup \mathcal{V}_b \cup \allowbreak  \mathcal{V}_c \cup \mathcal{V}_{nc}$. Note that each function node may provide multiple functions \cite{zhang2017network, chen2021optimal}. Thus, we assume a function node $v_i \in \mathcal{V}_{c}$ has a subset of the total functions $\mathcal{F}_i \subseteq \mathcal{F}$, where $\mathcal{F}_i := \{f_{i,n}\}_{n = 1, \ldots, N_i}$. Here, $f_{i, n}$ and $N_i$ denote the $n$-th function and the total number of functions for the function node $v_i$, respectively.

\begin{figure}[t!]
\centering
\includegraphics[scale=0.25]{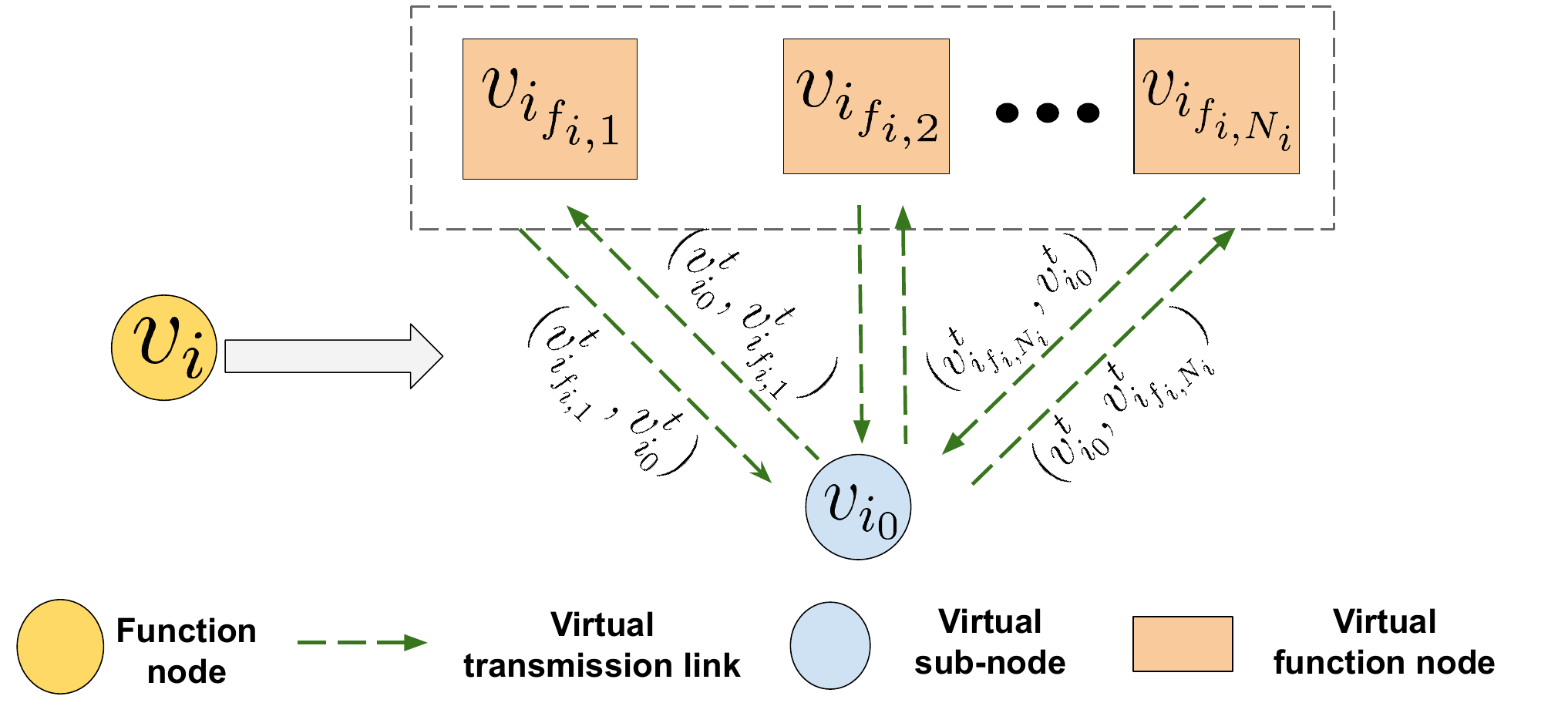}
\caption{Schematic of decoupling a function node into three types of components: virtual sub-node, virtual function nodes, and virtual transmission links.}
\label{fig:subnode_decompose}
\end{figure}

In the following subsections, we first explain how to construct the MT-TEG from the existing graph. Next, we introduce the detailed modeling process. Finally, we formulate a MILP to jointly optimize user association, the VNFs deployment, and flow routing strategy with the goal of maximizing the total processed data received by users.

\subsection{MF-TEG Model for the Time-Varying NFV-enabled SIN}
\label{subsec: MF-TEG Model for the Time-Varying NFV-enabled SIN}

Note that each function node $v_i \in \mathcal{V}_{c}$ can provide multiple functions for one task flow. Therefore, we adopt MF-TEG instead of traditional TEG to characterize this feature\cite{liu2023multi}. According to MF-TEG, as shown in Fig.~\ref{fig:subnode_decompose}, a function node $v_i \in \mathcal{V}_{c}$ can be decomposed into three types of components: a virtual sub-node $v_{i_0}$, $N_i$ virtual function nodes $\{v_{i_{f_{i,n}}}\}_{f_{i,n} \in \mathcal{F}_i}$, as well as virtual transmission links $\{(v_{i_0}, v_{i_{f_{i,n}}})\}_{{f_{i,n}} \in \mathcal{F}_i}$ and $\{(v_{i_{f_{i,n}}}, v_{i_0})\}_{{f_{i,n}} \in \mathcal{F}_i}$. Particularly, the virtual sub-node $v_{i_0}$ acts as a communication node without providing any function, while the node $v_{i_{f_{i,n}}}$ is the virtual function node decomposed by function node $v_i \in \mathcal{V}_c$. The links $(v_{i_0}, v_{i_{f_{i,n}}})$ and $(v_{i_{f_{i,n}}}, v_{i_0})$ are the virtual transmission links from the sub-node $v_{i_0}$ to the virtual function node $v_{i_{f_{i,n}}}$  and from the virtual function node $v_{i_{f_{i,n}}}$  to virtual sub-node $v_{i_0}$, respectively. We denote the set of the virtual sub-nodes and virtual function nodes decomposed by all function nodes as $\mathcal{V}_{vs}=\{v_{i_0}| v_i \in \mathcal{V}_c\}$ and $\mathcal{V}_{vc}=\{v_{i_{f_{i,n}}}| v_i \in \mathcal{V}_c,  f_{i,n} \in \mathcal{F}_i\}$, respectively. Besides, we denote the set of virtual function nodes decomposed by the function node $v_i \in \mathcal{V}_c$ as $\mathcal{V}_{vc, i}=\{v_{i_{f_{i,n}}}|  f_{i,n} \in \mathcal{F}_i\}$.

\begin{figure}[t!]
\centering
\includegraphics[scale=0.21]{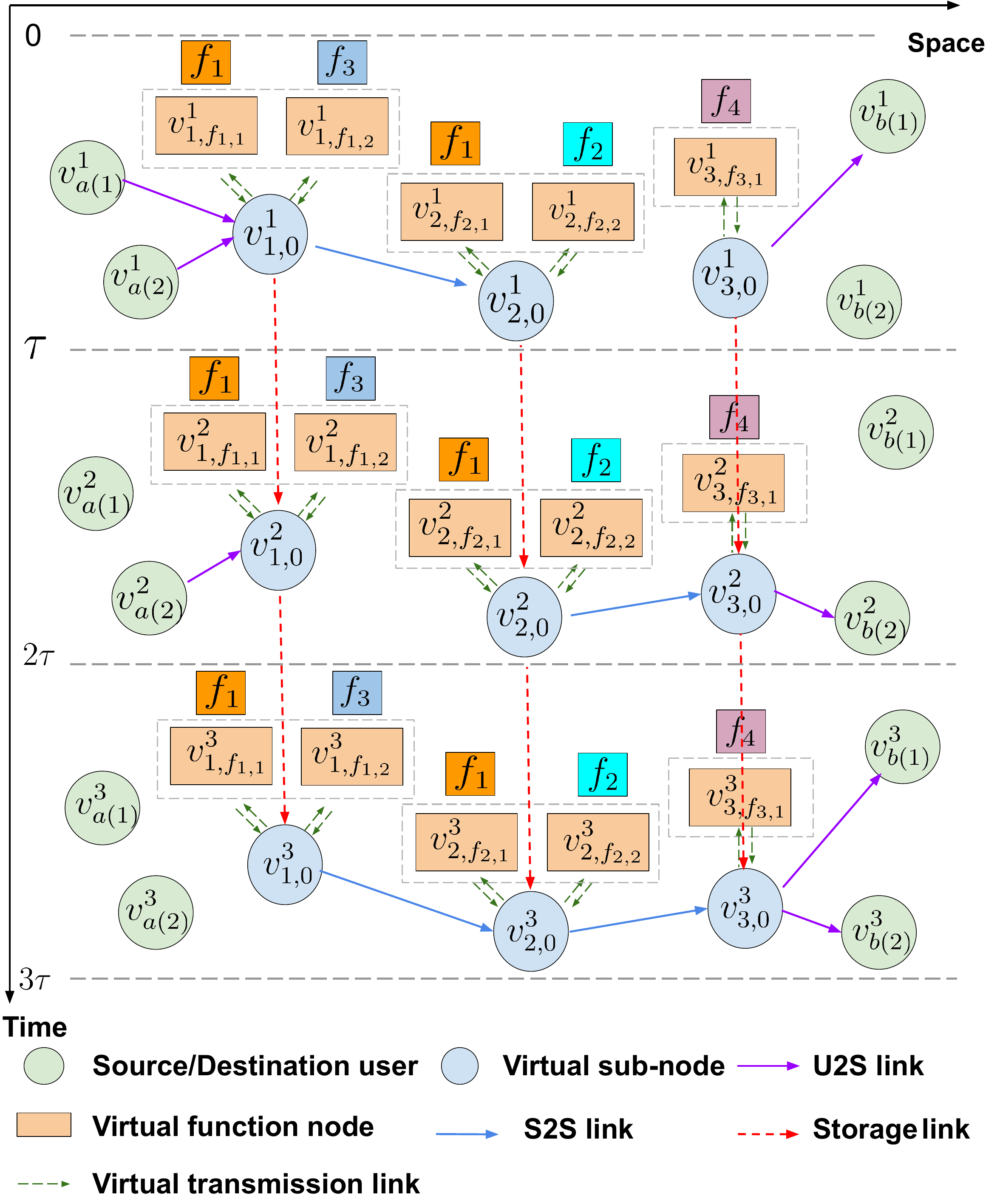}
\caption{MF-TEG for an example NFV-enabled SIN with three-time slots.}
\label{fig:MF-TEG}
\end{figure}

To simplify the exposition of the time-varying feature of this NFV-enabled SIN, we divide the time horizon $T$ into discrete time slots with slot duration $\delta$, which is indexed by $t \in \mathcal{T} :=\{1,\dots, T\}$. We assume that the network topology of the SIN is fixed during the time $t$, but may differ in different time slots. As illustrated in Fig.~\ref{fig:MF-TEG}, we model this dynamic network as an MF-TEG with $T$ layers. The NFV-enabled SIN can be further represented by $\mathcal{G}_M = (\mathcal{V}_M, \mathcal{A}_M)$, where $\mathcal{V}_M$ and $\mathcal{A}_M$ are the set of nodes and links in the MF-TEG, respectively. The components of the MF-TEG are detailed as follows: 

\emph{1). Nodes}: There are five kinds of nodes in the MF-TEG: source user nodes $\mathcal{V}_{M, a} = \{v_{a(l)}^t|  l \in \mathcal{L}, t \in \mathcal{T} \}$, destination user nodes $\mathcal{V}_{M, b} = \{v_{b(l)}^t|  l \in \mathcal{L}, t \in \mathcal{T} \}$, non-function nodes $\mathcal{V}_{M, nc} = \{v_i^t| v_i \in \mathcal{V}_{nc}, t \in \mathcal{T} \}$, virtual sub-nodes $\mathcal{V}_{M, vs} = \{v_{i_0}^t| v_{i_0} \in \mathcal{V}_{vs}, t \in \mathcal{T} \}$, and virtual function nodes $\mathcal{V}_{M, vc} = \{v_{i_{f_{i,n}}}^t| v_{i_{f_{i,n}}} \in \mathcal{V}_{vc}, t \in \mathcal{T} \}$. Therefore, we can represent the set of nodes in MF-TEG as $\mathcal{V}_M=\mathcal{V}_{M, a}\cup\mathcal{V}_{M, b}\cup\mathcal{V}_{M, nc}\cup\mathcal{V}_{M, vs}\cup\mathcal{V}_{M, vc}$.

\emph{2). Links}: There are three types of directed links in the MF-TEG: transmission links, virtual transmission links, and storage links \cite{liu2023multi}. The transmission links are denoted as $\mathcal{A}_{M, tr}=\{(v_i^t, v_j^t) | v_i, v_j \in \mathcal{V}, t \in \mathcal{T}\}$. The virtual transmission links can be described as $\mathcal{A}_{M, vl}= \mathcal{A}_{M, vl}^{\text{in}} \cup \mathcal{A}_{M, vl}^{\text{out}}$. Here, input virtual transmission link $\mathcal{A}_{M, vl}^{\text{in}} = \{(v_{i_0}^t, v_{i_{f_{i,n}}}^t)| v_{i_0} \in \mathcal{V}_{vs}, v_{i_{f_{i,n}}} \in \mathcal{V}_{vc,i}, t \in \mathcal{T}\}$ and output virtual transmission link $\mathcal{A}_{M, vl}^{\text{out}} = \{(v_{i_{f_{i,n}}}^t, v_{i_0}^t)| v_{i_0} \in \mathcal{V}_{vs}, v_{i_{f_{i,n}}} \in \mathcal{V}_{vc,i}, t \in \mathcal{T}\}$ are the sets of all virtual transmission links from the virtual sub-node to the virtual function node and from the virtual function node to the virtual sub-node, respectively. Besides, we denote the storage links as $\mathcal{A}_{M, st}=\{(v_i^t, v_i^{t+1}) | v_i \in \mathcal{V}_{vs}\cup\mathcal{V}_{nc}, t \in \mathcal{T}\}$, which represents the satellite node $i$ can store the data from the time $t$ to $t+1$. Hence, we can represent the set of links in MF-TEG as $\mathcal{A}_M=\mathcal{A}_{M, tr}\cup\mathcal{A}_{M,vl}\cup\mathcal{A}_{M,st}$. 

As the network topology of NFV-enabled SIN is dynamic, we denote the connectivity between nodes $v_i^t$ and $v_j^t$ at time $t$ by a binary indicator $K_{i,j}^t$, i.e., 
\begin{equation}
  K_{i, j}^t =
    \begin{cases}
      1 &  \text{if link}~(v_i^t,v_j^t) \in \mathcal{A}_{M,tr}~\text{is available at time $t$},\\
      0 & \text{otherwise}.
    \end{cases}       
\end{equation}
Note that we assume the virtual links $(v_i^t,v_j^t) \in \mathcal{A}_{M,vl}\cup\mathcal{A}_{M,st}$ are always available. Similar to \cite{yang2023space, yang2023group}, we introduce virtual flows, denoted as $\xi_{f_{l(k)}}$ ($ \forall f_{l(k)} \in \mathcal{F}(l)\cup \{f_{l(0)}\}$), to characterize the task flow with SFC constraints. Here, $\xi_{f_{l(k)}}$ represents the task flow $l$ that has just received the function $f_{l(k)}$, while virtual flow $f_{l(0)}$ represents the task flow that does not receive any function. Particularly, if a virtual flow $\xi_{f_{l(k-1)}}$ receives the function $f_{l(k)}$ from a virtual function node, the virtual flow will be changed to $\xi_{f_{l(k)}}$, otherwise, it is still $\xi_{f_{l(k-1)}}$. Note that each virtual function node is limited to offering a single function. For a virtual function node $ v_{i_{f_{i,n}}}^t \in \mathcal{V}_{M, vc}$, if it is capable of providing the specific function $f_{l(k)}$ required by the task flow $l$, then it can process the virtual flows $\xi_{f_{l(k-1)}}$. However, this virtual function node cannot process other virtual flow $ \xi_{f_{l(k^\prime)}} (\forall \xi_{f_{l(k^\prime)}} \neq \xi_{f_{l(k-1)}}, \xi_{f_{l(k^\prime)}}\in \mathcal{F}(l)\cup \{f_{l(0)}\})$. This constraint ensures that each node is dedicated to a specific function of the task flow. Moreover, if the virtual function node $v_{i_{f_{i,n}}}^t$ provides the function $f_{l(k)}$ for the task flow $l$, we assume the input virtual transmission link $(v_{i_0}^t, v_{i_{f_{i,n}}}^t)$ is only allowed to be used by the virtual flow $\xi_{f_{l(k-1)}}$. Consequently, other virtual flow  $ \xi_{f_{l(k^\prime)}} (\forall \xi_{f_{l(k^\prime)}} \neq \xi_{f_{l(k-1)}}, \xi_{f_{l(k^\prime)}} \in \mathcal{F}(l)\cup \{f_{l(0)}\})$ cannot use this link. Similarly, the output virtual transmission link $(v_{i_{f_{i,n}}}^t, v_{i_0}^t)$ is only allowed to be used by $\xi_{f_{l(k)}}$, while other virtual flow $\xi_{f_{l(k^\prime)}} (\forall \xi_{f_{l(k^\prime)}} \neq \xi_{f_{l(k)}}, \xi_{f_{l(k^\prime)}} \in \mathcal{F}(l) \cup \{f_{l(0)}\})$  is not allowed to use this link. 

In order to analyze the data flow in this NFV-enabled SIN, we denote $x_{i,j}^t(\xi_{f_{l(k)}})$,$ y_{i_0,i_{f_{i,n}}}^t(\xi_{f_{l(k)}})$, and $ z_{i_{f_{i,n}}, i_0}^t(\xi_{f_{l(k)}})$ as the cumulative amount of data transmitted on transmission link $(v_i^t, v_j^t) \in \mathcal{A}_{M, tr}$, virtual input transmission link $(v_{i_0}^t, v_{i_{f_{i,n}}}^t) \in \mathcal{A}_{M, vl}^{\text{in}}$, and virtual output transmission link $(v_{i_{f_{i,n}}}^t, v_{i_0}^t) \in \mathcal{A}_{M, vl}^{\text{out}}$ for the virtual flow $\xi_{f_{l(k)}}$ within time $t$, respectively. Moreover, we denote $o_i^{t, t+1}(\xi_{f_{l(k)}})$ as the total amount of data stored on storage link $(v_i^t, v_i^{t+1}) \in \mathcal{A}_{M, st}$ from time $t$ to $t+1$ for the virtual flow $\xi_{f_{l(k)}}$. We also utilize a binary indicator $H_{i_{f_{i,n}}, f^\prime}$ to describe whether virtual function node $v_{i_{f_{i,n}}} \in \mathcal{V}_{vc}$ provides function $f^\prime \in \mathcal{F}$, i.e., 
\begin{equation}
  H_{i_{f_{i,n}}, f^\prime} =
    \begin{cases}
      1 &  \text{if node}~v_{i_{f_{i,n}}}\in \mathcal{V}_{vc}~\text{provides function}~ f^\prime, \\
      0 & \text{otherwise}.
    \end{cases}       
\end{equation}

\subsection{User Association Model}
Since the network topology of NFV-enabled SIN is dynamic and the associations between users and satellites are changing over time, we utilize a binary variable $\phi^{\text{U2S},t}_{a(l), i}$ to represent the association between source user $v_{a(l)}$ and satellite node $ v_i \in \mathcal{V}_s$ at time $t$, where $\phi^{\text{U2S},t}_{a(l), i} = 1$ indicates source user $v_{a(l)}$ is associated with satellite node $ v_i \in \mathcal{V}_s$ at time $t$, and otherwise $\phi^{\text{U2S},t}_{a(l), i} = 0$. Similarly, a binary variable $\phi^{\text{S2U},t}_{i, b(l)}$ is used to describe the association between satellite $v_i \in \mathcal{V}_s$ and destination user $v_{b(l)}$ at time $t$, where $\phi^{\text{S2U},t}_{i, b(l)} = 1$ indicates satellite $v_i \in \mathcal{V}_s$ is associated with destination user $v_{b(l)}$ at time $t$, and otherwise $\phi^{\text{S2U},t}_{i, b(l)} = 0$. Since the users can only connect to the satellite when the link is available, we have
\begin{align} 
\quad & \phi^{\text{U2S},t}_{a(l),i} \leq K_{a(l),i}^t, \quad \forall l \in \mathcal{L}, t \in \mathcal{T}, i \in \mathcal{V}_s  \label{cons: ur_a_asso_1},\\
\quad & \phi^{\text{S2U},t}_{i,b(l)} \leq K_{i,b(l)}^t, \quad \forall l \in \mathcal{L}, t \in \mathcal{T}, i \in \mathcal{V}_s \label{cons: ur_a_asso_2}.
\end{align}
Due to the antenna limitation of terrestrial users, each user can connect to only one satellite at time $t$, i.e.,  
\begin{align}
    \sum_{i \in \mathcal{V}_{s}}\phi^{\text{U2S},t}_{a(l), i} = 1, \quad \forall l \in \mathcal{L}, t \in \mathcal{T},\\
    \sum_{i \in \mathcal{V}_{s}} \phi^{\text{S2U},t}_{i, b(l)} = 1, \quad \forall l \in \mathcal{L}, t \in \mathcal{T}. 
\end{align}

Similarly, we assume that each satellite $v_i \in \mathcal{V}_s$ can connect to at most $C_{i}^{\text{U2S}}$ source users and $C_{i}^{\text{S2U}}$ destination users in time $t$, respectively, i.e.,
\begin{align}
&\sum_{l \in \mathcal{L}}\phi^{\text{U2S},t}_{a(l),i} \leq C_{i}^{\text{U2S}}, \quad \forall i \in \mathcal{V}_{s}, t \in \mathcal{T} \label{cons: ur_b_asso_1},\\
&\sum_{l \in \mathcal{L}} \phi^{\text{S2U},t}_{i, b(l)} \leq C_{i}^{\text{S2U}}, \quad \forall  i \in \mathcal{V}_{s}, t \in \mathcal{T}. \label{cons: ur_b_asso_2}
\end{align}

\subsection{VNF Deployment Model} 
We introduce the variables $\lambda_{i_{f_{i,n}}, f_{l(k)}} \in \{0, 1\}$ to denote whether the virtual function node $v_{i_{f_{i,n}}}$ serves function $f_{l(k)}$ for the task flow $l$, where $\lambda_{i_{f_{i,n}}, f_{l(k)}}  = 1$ indicates the virtual function node $v_{i_{f_{i,n}}}$ serves the function $f_{l(k)}$ for flow $l$, and otherwise $\lambda_{i_{f_{i,n}}, f_{l(k)}}  = 0$. Note that this virtual function node $v_{i_{f_{i,n}}}$ cannot serve the function $f_{l(k)}$ for the task flow $l$ if it does not carry the corresponding function $f_{l(k)}^\prime \in \mathcal{F}$. Therefore, we have
\begin{align}
    \lambda_{i_{f_{i,n}}, f_{l(k)}} \leq H_{i_{f_{i,n}}, f^\prime_{l(k)}}, \quad &\forall i_{f_{i,n}} \in \mathcal{V}_{vc},  f_{l(k)} \in \mathcal{F}(l), l \in \mathcal{L}.
\end{align}
Similar to \cite{zhang2017network, chen2021optimal, yang2023space}, we assume that each function $f_{l(k)} \in \mathcal{F}(l)$ is processed by only one virtual function node to reduce the coordination overhead, which can be expressed as 
\begin{align}
    \sum_{v_{i_{f_{i,n}}} \in \mathcal{V}_{vc}} \lambda_{i_{f_{i,n}}, f_{l(k)}} = 1, \quad \forall f_{l(k)} \in F(l), l \in \mathcal{L}.
\end{align}

\subsection{Communication Model}
 In the SIN,  there are three types of transmission links: U2S transmission links, S2S transmission links, and S2U transmission links. Here, U2S transmission links are denoted as $\mathcal{A}^{\text{S2U}}=\{(v_i^t, v_j^t) | v_i \in \mathcal{V}_{a}, v_j \in \mathcal{V}_s, t \in \mathcal{T}\}$, S2S transmission links are denoted as $\mathcal{A}^{\text{S2S}}=\{(v_i^t, v_j^t) | v_i \in \mathcal{V}_{s}, v_j \in \mathcal{V}_s, t \in \mathcal{T}\}$, and S2U transmission links are denoted as $\mathcal{A}^{\text{S2U}}=\{(v_i^t, v_j^t) | v_i \in \mathcal{V}_{s}, v_j \in \mathcal{V}_b, t \in \mathcal{T}\}$. Therefore, the transmission links can be described as  $\mathcal{A}_{M,tr}=\mathcal{A}^{\text{S2S}} \cup \mathcal{A}^{\text{S2U}} \cup \mathcal{A}^{\text{U2S}}$. To elaborate on the channel models between node $v_i^t$ and $v_j^t$ ($v_i^t, v_j^t \in \mathcal{V}_{M}$), we start with the notations. $P_{i,j}^t$ is the transmission power from nodes $v_i^t$ to $v_j^t$. $G_{i,j}^{\text{TR},t}, G_{i,j}^{\text{RE},t}$ are the transmission antenna gain of node $v_i^t$ and receive antenna gain of node $v_j^t$, respectively. $L_l$ is the total line loss. $L_{i,j}^t = (\frac{c}{4\pi d_{i,j}^t \nu_{i,j}^t})^2$ is the free space loss, in which $c$ is the speed of light, $d_{i,j}^t$ is the maximum slant range, and $\nu_{i,j}^t$ is the center frequency. Then, we can utilize the following models to characterize them.

\emph{1) S2S link:} According to \cite{golkar2015federated}, the achievable data transmission rate of S2S link can be expressed as
\begin{equation}
    R(v_i^t, v_j^t) = \frac{P_{i,j}^tG_{i,j}^{\text{TR},t}G_{i,j}^{\text{RE},t}L_{i,j}^tL_l}{k_BT_sM(E_b/N_0)},  \forall (v_i^t, v_j^t) \in \mathcal{A}^{\text{S2S}},
\end{equation}
where $k_B$ is the Boltzmann's constant. $T_s$ is the system noise temperature. $M$ is the link margin. $E_b/N_0$ is the ratio of received energy-per-bit to noise-density.

\emph{2) U2S/S2U link:} Based on the Shannon formula, the achievable data transmission rate for U2S and S2U links between satellites and users can be calculated as
\begin{equation}
     R(v_i^t, v_j^t) =  B_{i,j}\log_2(1+\text{SNR}_{i,j}^t), \forall (v_i^t, v_j^t) \in \mathcal{A}^{\text{U2S}} \cup \mathcal{A}^{\text{S2U}},
\end{equation}
where $B_{i,j}$ is the available bandwidth between the satellites and users. Besides, $\text{SNR}_{i,j}$ is expressed as
\begin{equation}
     \text{SNR}_{i,j}^t =  \frac{P_{i,j}^tG_{i,j}^{\text{TR},t}G_{i,j}^{\text{RE},t}L_{i,j}^tL_l}{N_0}, \forall (v_i^t, v_j^t) \in \mathcal{A}^{\text{U2S}} \cup \mathcal{A}^{\text{S2U}},
\end{equation}
where $N_0$ is the additional Gaussian white noise power.

Therefore, the maximum amount of data that can be transmitted on the link $(v_i^t, v_j^t) \in \mathcal{A}_{M, tr} $ within time $t$ can be written as
\begin{equation}
\label{eq:revised_12}
    C(v_i^t, v_j^t) = R(v_i^t, v_j^t)\cdot \delta, \quad \forall (v_i^t, v_j^t) \in \mathcal{A}_{M,tr}, t \in \mathcal{T},
\end{equation}
where $\delta$ is the duration of the time $t$.

Since the data amount on the U2S link $(v_{a(l)}^t, v_i^t) \in \mathcal{A}_{M, tr}$ is restricted on both the channel capacity $C(v_i^t, v_j^t)$ and the association between user and satellite $\phi_{a(l),i}^{\text{U2S},t}$, the following constraint needs to be satisfied:
\begin{align}
    x_{a(l),i}^{t}\left(\xi_{f_{l(0)}}\right)  \leq  & \phi_{a(l),i}^{\text{U2S},t} C\left(v_{a(l)}^t, v_i^t\right), \quad \forall l \in \mathcal{L}, i \in \mathcal{V}_{s}, t \in \mathcal{T}.
\end{align}
Similarly, we have a constraint on the S2U link  $(v_i^t, v_{b(l)}^t) \in \mathcal{A}_{M, tr}$ as
\begin{align}
    x_{i,b(l)}^{t}\left(\xi_{f_{l(K)}}\right)  \leq  & \phi_{i,b(l)}^{\text{S2U},t} C\left(v_i^t, v_{b(l)}^t\right), \quad \forall l \in \mathcal{L}, i \in \mathcal{V}_{s}, t \in \mathcal{T}.
\end{align}
For the S2S transmission link $(v_i, v_j) \in \mathcal{A}_{M, tr}$, we have the following transmission link capacity constraints:
\begin{align}
\sum_{l \in \mathcal{L}} \sum_{f_{l(k)} \in\mathcal{F}(l)\cup\{f_{l(0)}\}}\,x_{i,j}^{t}\left(\xi_{f_{l(k)}}\right) \leq &  K_{i,j}^tC\left(v_i^{t},v_j^{\,t}\right), \notag \\
& \forall i, j \in  \mathcal{V}_s, t \in \mathcal{T}.  
\end{align}

Unlike physical transmission links, we assume the channel capacity of the input transmission link $(v_{i_0}^t, v_{i_{f_{i,n}}}^t) \in \mathcal{A}_{M, vl}^{\text{in}}$ and the output virtual transmission link $(v_{i_{f_{i,n}}}^t, v_{i_0}^t) \in \mathcal{A}_{M, vl}^{\text{out}}$ are constants, denoted by $C(v_{i_0}^t, v_{i_{f_{i,n}}}^t)$ and $C(v_{i_{f_{i,n}}}^t, v_{i_0}^t)$, respectively. The total amount of data transmitted over the virtual transmission link $(v_{i_0}^t, v_{i_{f_{i,n}}}^t)$ is constrained by its channel capacity $C(v_{i_0}^t, v_{i_{f_{i,n}}}^t)$ and whether the virtual function node $v_{i_{f_{i,n}}}^t$ serves the function $f_{l(k-1)}$ for task flow $l$, which can be described by
\begin{align}
 &y_{i_0,i_{f_{i,n}}}^{t} \left(\xi_{f_{l(k-1)}}\right) \leq   \lambda_{i_{f_{i,n}},f_{l(k)}} C\left(v_{i_0}^t, v_{i_{f_{i,n}}}^t \right), \quad \notag\\
 &\forall l\in\mathcal{L}, f_{l(k-1)}\in{\mathcal{F}}(l)\cup\{f_{l(0)}\},(v_{i_0}^t, v_{i_{f_{i,n}}}^t)\in{\mathcal{A}}_{M,vl}^{\text{in}}.
\end{align}
Similarly, for the virtual transmission link $(v_{i_{f_{i,n}}}^t, v_{i_0}^t)$, we have the following channel capacity constraint: 
\begin{align}
 z_{i_{f_{i,n}},i_0}^{t}&\left(\xi_{f_{l(k)}}\right)  \leq \lambda_{i_{f_{i,n}},f_{l(k)}}{ C}\left(v_{i_{f_{i,n}}}^t, v_{i_0}^t\right), \notag\\
 & \forall l\in\mathcal{L},f_{l(k)}\in{\mathcal{F}}(l),(v_{i_{f_{i,n}}}^t, v_{i_0}^t) \in{\mathcal{A}}_{M,vl}^{\text{out}}.
\end{align}

\subsection{Computation Model}
We denote the computation capacity of function node $v_i \in \mathcal{V}_c$ as $C_i^t$ in time $t$. Since all virtual function nodes $v_{i_{f_{i,n}}}^t$ decomposed by the function node $v_i$ consume the total computation capacity of function node $v_i$ at time $t$, we have the following computation constraint:
\begin{align}
 \sum_{l \in \mathcal{L}}\sum_{f_{l(k) \in \mathcal{F}(l)}}\sum_{v_{i_{f_{i,n}}}^t \in \mathcal{V}_{M, vc, i}} \kappa_{f_{l(k)}} y_{i_0,i_{f_{i,n}}}^{t} &\left(\xi_{f_{l(k-1)}}\right) \leq C_i^t, \notag\\
 &\forall v_{i_0}^t \in \mathcal{V}_{M, vs}, 
\end{align}
where $\kappa_{f_{l(k)}}$ is the computation factor for processing function $f_{l(k)}$ and $\mathcal{V}_{M, vc, i} = \{v_{i_{f_{i,n}}} | f_{i,n} \in \mathcal{F}_i, t \in \mathcal{T}\}$ is the set of virtual function nodes decomposed by function node $i \in \mathcal{V}_c$ at the time $t$.

\subsection{Storage Model}
We denote the maximum data amount that can be stored on the storage link $(v_i^t, v_i^{t+1})$ as $C(v_i^t, v_i^{t+1})$. Consequently, we have the following storage capacity constraint:
\begin{align}
\sum_{l \in \mathcal{L}}\sum_{f_{l(k)} \in \mathcal{F}(l) \cup \{f_{l(0)}\}}o_i^{t, t+1}(\xi_{f_{l(k)}}) \leq C(v_i^t, v_i^{t+1}), \notag\\
\forall (v_i^t, v_i^{t+1}) \in \mathcal{A}_{M, st}.
\end{align}

\subsection{Data Flow Model}
Since a non-function node $v_i^t \in \mathcal{V}_{M, nc}$ just has the communication capacity, the total data amount of input virtual flow $\xi_{f_{l(k)}}$ should be equal to the total data amount of output virtual flow $\xi_{f_{l(k)}}$, which can be written as
\begin{flalign}
&\sum_{v_j^{t}:(v_j^{t},v_i^{t})\in {\mathcal{A}}_{M,tr}} x_{j,i}^{t}\left(\xi_{f_{l(k)}}\right)+o_{i}^{t-1,t}\left(\xi_{f_{l(k)}}\right) \notag \\
& = \sum_{v_j^{t}:(v_i^{t},v_j^{t})\in {\mathcal{A}}_{M,tr}} x_{i,j}^{t}\left(\xi_{f_{l(k)}}\right)+o_{i}^{t, t+1}\left(\xi_{f_{l(k)}}\right), \notag \\
& \quad \quad \forall v_i^{t}\in \mathcal{V}_{M,nc}, l\in\mathcal{L},f_{l(k)}\in{\mathcal{F}}(l)\cup\{f_{l(0)}\}.
\end{flalign}
Similarly, we have the following constraint for the sub-node $i_0^t \in \mathcal{V}_{M, vs}$:
\begin{align}
&\sum_{v_{i_{f_{i,n}}}^{t} \in \mathcal{V}_{M, vc, i}} z_{i_{f_{i,n}}, i_0}^{t}\left(\xi_{f_{l(k)}}\right) + \sum_{v_j^{t}:(v_j^{t},v_{i_0}^{t})\in {\mathcal{A}}_{M,tr}} x_{j,i_0}^{t}\left(\xi_{f_{l(k)}}\right) \notag \\
& +o_{i_0}^{t-1,t}\left(\xi_{f_{l(k)}}\right)  = \sum_{v_{i_{f_{i,n}}}^{t} \in \mathcal{V}_{M, vc, i}} y_{i_0,i_{f_{i,n}}}^{t}\left(\xi_{f_{l(k)}}\right)  \notag \\
&+ \sum_{v_j^{t}:(v_{i_0}^{t},v_j^{t})\in {\mathcal{A}}_{M,tr}} x_{i_0,j}^{t}+ o_{i_0}^{t, t+1}\left(\xi_{f_{l(k)}}\right), \notag \\
&\quad \quad \forall v_{i_0}^{t}\in \mathcal{V}_{M,vs}, l \in \mathcal{L}, f_{l(k)}\in{\mathcal{F}}(l) \cup \{f_{l(0)}\}.
\end{align}

Moreover, for the virtual function node $v_{i_{f_{i,n}}}^t \in \mathcal{V}_{M, vc}$, if $v_{i_{f_{i,n}}}^t$ serves the function $f_{l(k)}$ for the flow $l$, the virtual flow $\xi_{f_{l(k-1)}}$ is processed by the virtual function node $v_{i_{f_{i,n}}}^t$ and transformed into a new virtual flow $\xi_{f_{l(k)}}$. We denote a scaling factor $\beta_{\xi_{f_{l(k-1)}}, \xi_{f_{l(k)}}}$ that represents the ratio between the data amounts of $\xi_{f_{l(k-1)}}$ and $\xi_{f_{l(k)}}$. Then, we have
\begin{align}
\label{eq:compute_ratio}
    &y_{i_0,i_{f_{i,n}}}^{t} \left(\xi_{f_{l(k-1)}}\right) = 	\beta_{\xi_{f_{l(k-1)}}, \xi_{f_{l(k)}}}z_{i_{f_{i,n}}, i_0}^{t}\left(\xi_{f_{l(k)}}\right), \notag\\
    & \quad \quad   \forall l\in\mathcal{L},f_{l(k-1)}\in{\mathcal{F}}(l)\cup\{f_{l(0)}\}, v_{i_0, i_{f_{i,n}}}^t \in \mathcal{V}_{M, vc}.
\end{align}

Note that the output data of the source user node $v_{a(l)}^t \in \mathcal{V}_{M,a}$ is only the virtual flow $\xi_{f_{l(0)}}$, and other virtual flow $\xi_{f_{l(k)}}(\forall f_{l(k)} \neq f_{l(0)} )$  should be zero, which can be expressed as 
\begin{align}
x_{a(l), i}^t \left(\xi_{f_{l(k)}}\right) = 0, \quad \forall & l \in \mathcal{L}, f_{l(k)} \in \mathcal{F}(l), \notag \\
&(v_{a(l)}^t, v_i^t) \in \mathcal{A}_{M, tr}.
\end{align}
Similarly, the input data of the destination user $v_{b(l)}^t \in \mathcal{V}_{M,b}$ is only the virtual flow $\xi_{f_{l(K)}}$, and other virtual flow $\xi_{f_{l(k)}}(\forall f_{l(k)} \neq f_{l(K)} )$  should be zero, which is described as 
\begin{align}
x_{i, b(l)}^t \left(\xi_{f_{l(k)}}\right) = 0, \quad \forall & l \in \mathcal{L}, f_{l(k)} \neq f_{l(K)} \in \mathcal{F}(l) \cup \{f_{l(0)}\}, \notag \\
&(v_i^t, v_{b(l)}^t) \in \mathcal{A}_{M, tr} \label{const: data_flow_5}.
\end{align}

Finally, the total processed data that successfully reaches the destination users is expressed as 
\begin{align}
Q = \sum_{t \in \mathcal{T}}\sum_{l \in \mathcal{L}} \sum_{i \in \mathcal{V}_s}  x_{i,b(l)}^{t}\left(\xi_{f_{l(K)}}\right).
\end{align}

\subsection{Problem Formulation}
\label{subsec: problem_formulation}
To maximize the total processed data that successfully reaches the destination users, we jointly optimize the user associations $\bm{\phi}=\{\phi^{\text{U2S},t}_{a(l),j},\phi^{\text{S2U},t}_{i,b(l)}\}$, function service $\bm{\lambda}=\{\lambda_{i_{f_{i,n}}, f_{l(k)}}\}$, cumulative amount of data $\mathbf{x}=\{x_{i,j}^t(\xi_{f_{l(k)}})\}$ transmitted on transmission link $(i^t, j^t) \in \mathcal{A}_{M, s2s}$, cumulative amount of data $\mathbf{y}=\{y_{i_0,i_{f_{i,n}}}^t(\xi_{f_{l(k)}})\}$ transmitted on virtual transmission link $(i_0^t, i_{f_{i,n}}^t) \in \mathcal{A}_{M, vl}^{\text{in}}$, cumulative amount of data $\mathbf{z}=\{z_{i_{f_{i,n}}, i_0}^t(\xi_{f_{l(k)}})\}$ transmitted on virtual transmission link  $(i_{f_{i,n}}, i_0^t) \in \mathcal{A}_{M, vl}^{\text{out}}$, and cumulative amount of data $\mathbf{o}=\{o_{i}^{t, t+1}(\xi_{f_{l(k)}})\}$ on storage link $(i^t, i^{t+1}) \in \mathcal{A}_{M, st}$, i.e., 
\begin{subequations}
\label{P0}
\begin{align}
 \textbf{P}_0: \max_{\bm{\phi, \lambda}, \mathbf{x, y, z, o}} \quad &  Q\\
\text{s.t.}
\quad & \mathbf{ x, y, z, o} \geq 0, \\
\quad & \bm{\phi, \lambda} \in \{0,1\},\\
\quad &\eqref{cons: ur_a_asso_1}- \eqref{const: data_flow_5}.
\end{align}
\end{subequations}
For ease of presentation, we rewrite problem $\textbf{P}_0$ in an abstract formulation as follows:
\begin{subequations} 
\label{P1}
\begin{align}
 \textbf{P}_1: \min_{\mathbf{q, w}} \quad &  -\mathbf{c}^\intercal\mathbf{q} \\
\text{s.t.}
\quad & \mathbf{B}_1\mathbf{q}+ \mathbf{G}_1\mathbf{w} \leq \mathbf{d}_1, \\
\quad & \mathbf{B}_2\mathbf{q}+ \mathbf{G}_2\mathbf{w} = \mathbf{d}_2, \\
\quad & \mathbf{G}_3\mathbf{w} \leq \mathbf{d}_3, \\
\quad & \mathbf{q} \geq 0,\\
\quad & \mathbf{w} \in \{0,1\},
\end{align}
\end{subequations}
where the symbol $\mathbf{q}=\left(\mathbf{x,y,z,o}\right)$, and the symbol $\mathbf{y}=\left(\bm{\phi,\lambda}\right)$. $\mathbf{q}$ is a continuous variable vector, and $\{\mathbf{B}_n\}_{n = 1,2,3}$ is its corresponding coefficient matrix in constraints. $\mathbf{w}$ is a binary variable vector, and $\{\mathbf{G}_n\}_{n = 1,2,3}$ is its corresponding coefficient matrix in constraints. $\mathbf{c}^\intercal$ is the coefficient vector of variable $\mathbf{q}$ in the objective function. $\{\mathbf{d}_n\}_{n = 1,2,3}$ is the corresponding right-hand side constant vector. By solving this problem, we can obtain the optimal user association, VNFs deployment, and flow routing strategy. However, as we can see, the continuous decision variables $\mathbf{q}$ are tightly coupled with binary decision variables $\mathbf{w}$, which makes problem $\textbf{P}_1$ a large-scale MILP that is challenging to solve\footnote{Note that the complexity of problem $\textbf{P}_1$ can be written as $\mathcal{O}(2^{n_0}\cdot m_0^{3.5})$, where $n_0$ is the total number of binary decision variables and $m_0$ is the total number of continuous decision variables in problem $\textbf{P}_1$.}.

\section{Solution Methodology} 
\label{sec: solution}
In this section, we focus on designing an efficient algorithm for the central network controller to solve problem $\textbf{P}_1$, which jointly optimizes user association, VNFs deployment, and flow routing strategy (U-VNF-R). First, we propose an innovative hybrid quantum-classical algorithm to address $\textbf{P}_1$ in Subsection \ref{HQCBD}. Subsequently, we further design a multi-cut strategy to enhance the convergence speed of the proposed algorithm in Subsection \ref{MC}.

\subsection{Hybrid Quantum-classical Benders' Decomposition}
\label{HQCBD}
Problem $\textbf{P}_1$ is hard to solve since the continuous decision variable $\mathbf{p}$ is coupled with the integer decision variable $\mathbf{w}$. We first leverage the BD to decompose the problem $\textbf{P}_1$ into a master problem and a subproblem. The master problem only involves the binary optimization decision variables, which can be reformulated to the QUBO formulation and solved by a quantum computer. On the other hand, the subproblem only involves the continuous optimization decision variables, which can be solved by a classical computer. The solutions of the master problem and subproblem provide the performance lower and upper bounds of the problem $\textbf{P}_1$, respectively. We iteratively solve these two problems until the gap between the upper and lower bound is sufficiently small. We detail the subproblem and master problem as follows.

\subsubsection{Classical Optimization for Subproblem}
For the given binary variables $\mathbf{w}^{(\zeta)}$ generated by the master problem at the $(\zeta-1)$-th iteration, the subproblem can be written as
\begin{subequations} 
\begin{align}
 \textbf{SP}_1: \min_{\mathbf{q}} \quad &  -\mathbf{c}^\intercal\mathbf{q} \\
\text{s.t.}
\quad & \mathbf{B}^*\mathbf{q} \leq \mathbf{d}^* - \mathbf{G}^*\mathbf{w}^{(\zeta)}, \\
\quad & \mathbf{q} \geq 0,\\
\quad & \mathbf{B}^*=[\mathbf{B}_1, \mathbf{B}_2]^\intercal,\quad \mathbf{d}^*=[\mathbf{d}_1, \mathbf{d}_2]^\intercal, \\
\quad & \mathbf{G}^*=[\mathbf{G}_1, \mathbf{G}_2]^\intercal.
\end{align}
\end{subequations}
According to LP’s duality, the dual problem of subproblem $\textbf{SP}_1$ can be expressed as
\begin{subequations} \label{sp_p1} 
\begin{align}
 \textbf{SP}_2: \max_{\mathbf{u}} \quad &   (\mathbf{d}^* -\mathbf{G}^*\mathbf{w}^{(\zeta)})^\intercal \mathbf{u} \\
\text{s.t.}
\quad & ({\mathbf{B}^*})^\intercal\mathbf{u} \leq  -\mathbf{c}, \label{sp_p1_1}\\
\quad & \mathbf{u} \geq  0, \label{sp_p1_2}\\
\quad & \mathbf{B}^*=[\mathbf{B}_1, \mathbf{B}_2]^\intercal, \quad
 \mathbf{d}^*=[\mathbf{d}_1, \mathbf{d}_2]^\intercal, \\
\quad & \mathbf{G}^*=[\mathbf{G}_1, \mathbf{G}_2]^\intercal,
\end{align}
\end{subequations}
where $\mathbf{u}$ is the dual variable. This problem can be efficiently solved by classical LP numerical solvers such as Gurobi \cite{Gurobi}. We also check the feasibility of the subproblem $\textbf{SP}_2$. Specifically, if the inner product between $ (\mathbf{d}^* -\mathbf{G}^*\mathbf{w}^{(\zeta)})^\intercal$ and its extreme ray $e_1$ is negative, the subproblem is infeasible. Then, we can generate a new feasibility cut $C^{\text{F}}$, i.e., 
\begin{equation}
    (\mathbf{d}^* -\mathbf{G}^*\mathbf{w}^{(\zeta)})^\intercal e_1 \leq 0.
\end{equation}
On the other hand, if the subproblem is feasible, we yield a new optimality cut $C^{\text{O}}$ by its extreme point $e_2$, i.e., 
\begin{equation}
    (\mathbf{d}^* -\mathbf{G}^*\mathbf{w}^{(\zeta)})^\intercal e_2 \leq \theta,
\end{equation}
where $\theta$ is the optimal value of subproblem $\textbf{SP}_2$.  Both the feasibility cuts and optimality cuts generated in the current iteration are used to guide the future direction of the master problem in the next iteration.

\subsubsection{Quantum Optimization for Master Problem} 
After solving subproblem $\textbf{SP}_2$, we can obtain a new extreme ray $e_1^{(\zeta)}$ or point $e_2^{(\zeta)}$ for feasibility cut or optimality cut at the $\zeta$-th iteration, respectively. Moreover, we define $\mathcal{E}_1$ and $\mathcal{E}_2$ as the sets of all known iteration indices associated with the subproblem being infeasible and feasible, respectively. Then, we can formulate the master as 
\begin{subequations} 
\label{MP1}
\begin{align}
 \textbf{MP}_1: \min_{\mathbf{w}, \theta} \quad &  \theta \\
\text{s.t.}
\quad & \mathbf{G}_3\mathbf{w} \leq \mathbf{d}_3, \label{MP_1: original}\\
\quad &  (\mathbf{d}^* -\mathbf{G}^*\mathbf{w})^\intercal e_1^{(\zeta)} \leq 0, \quad \forall \zeta \in \mathcal{E}_1, \label{MP_1: feasiable}\\
\quad & (\mathbf{d}^* -\mathbf{G}^*\mathbf{w})^\intercal e_2^{(\zeta)} \leq \theta, \quad \forall \zeta \in \mathcal{E}_2, \label{MP_1: optimal}\\
\quad & \mathbf{w} \in \{0,1\}.
\end{align}
\end{subequations}
 By adding the feasibility cuts and optimality cuts, the search region for the globally optimal solution is gradually reduced \cite{rahmaniani2017benders}. Besides, the objective value of problem $\textbf{MP}_1$ is the performance lower bound of problem $\textbf{P}_1$. 

Although the scale of master problem $\textbf{MP}_1$ is much smaller than the original problem $\textbf{P}_1$, it is still a MILP, which is hard for classical computers to tackle. Moreover, QA only accepts the QUBO formulation as input. Unlike the QUBO formulation, the master problem $\textbf{MP}_1$ has a continuous variable as the objective function and contains multiple constraints. Therefore, we utilize the following steps to convert it into the QUBO formulation so that it can be solved by QA \cite{zhao2022hybrid}. 

\textbf{Objective function reformulation:} Since the QA only accepts a quadratic polynomial over binary variables, we first approximate the continuous variable $\theta$ using a binary vector $\mathbf{v}$ with the length of $U$ bits, i.e., 
\begin{align}
    \Bar{\theta} =& \sum_{i=-\underline{u}_+}^{\overline{u}_+}v_{(i+\underline{u}_+)}2^iv_{(i+\underline{u}_+)}- \nonumber\\ &\sum_{j=0}^{\overline{u}_-}w_{(j+1+\underline{u}_++\overline{u}_+)}2^jv_{(j+1+\underline{u}_++\overline{u}_+)} = \Bar{\theta}(\mathbf{v}),
\end{align}
where $U=1+\underline{u}_++\overline{u}_++\overline{u}_-$. Here, $\overline{u}_-, \overline{u}_+, \underline{u}_+$ represent the number of bits that are assigned to represent the negative integer, positive integer, and positive decimal part of $\theta$, respectively. 

\textbf{Constraints reformulation:} 
After reformulating the objective function in the master problem $\textbf{MP}_1$, we can obtain a constraint ILP master problem, which is still not the QUBO formulation. Thus, it is necessary to further reformulate the constrained ILP master problem into an unconstrained QUBO using penalty terms.According to the constraint-penalty pair principle in \cite{zhao2022hybrid}, we convert the constraints \eqref{MP_1: original}, \eqref{MP_1: feasiable}, and \eqref{MP_1: optimal} as follows, respectively,
\begin{align}
    f_Q^{\eqref{MP_1: original}}(\mathbf{w}) = 	\eta_{1}(\mathbf{G}_3\mathbf{w} - \mathbf{d}_3 + \sum_{\psi=0}^{\overline{\psi}_{1}}2^\psi{\gamma}_{1,\psi})^2,
\end{align}
where $\overline{\psi}_{1} = \big\lceil \log_2\big(\min_{\mathbf{w}}(
     \mathbf{d}_3 -\mathbf{G}_3\mathbf{w})\big) \big\rceil$,
\begin{flalign}
    f_Q^{\eqref{MP_1: feasiable}}(\mathbf{w}) = 
\eta_{2}^{(\zeta)}\Biggl({\mathbf{d}^*}^\intercal e_1^{(\zeta)}-(e_1^{(\zeta)})^\intercal\mathbf{G}^*\mathbf{w} + &\sum_{\psi=0}^{\overline{\psi}_{2}^{(\zeta)}}2^\psi{\gamma}_{2,\psi}^{(\zeta)}\Biggl)^2, \notag \\
& \forall \zeta \in \mathcal{E}_1,
\end{flalign}
where $\overline{\psi}_{2}^{(\zeta)} = \big\lceil \log_2\big(\min_{\mathbf{w}}((e_1^{(\zeta)})^\intercal\mathbf{G}^*\mathbf{w}-{\mathbf{d}^*}^\intercal e_1^{(\zeta)}\big) \big\rceil$, and
\begin{flalign}
    f_Q^{\eqref{MP_1: feasiable}}(\mathbf{w,v}) =  \eta_{3}^{(\zeta)}\Biggl(&{\mathbf{d}^*}^\intercal e_1^{(\zeta)} -(e_1^{(\zeta)})^\intercal\mathbf{G}^*\mathbf{w} -\Bar{\theta}(\mathbf{v}) \notag\\
& +\sum_{\psi=0}^{\overline{\psi}_{2}^{(\zeta)}}2^\psi{\gamma}_{2,\psi}^{(\zeta)}\Biggl)^2, \forall \zeta \in \mathcal{E}_2,
\end{flalign}
where $\overline{\psi}_{3}^{(\zeta)} = \big\lceil \log_2\big(\min_{\mathbf{w, v}}((e_1^{(\zeta)})^\intercal\mathbf{G}^*\mathbf{w}-{\mathbf{d}^*}^\intercal e_1^{(\zeta)}+\Bar{\theta}(\mathbf{v})\big) \big\rceil$.

Here, $\mathbf{r}$ is the binary slack variables, $\Bar{\bm{\psi}}$ is the upper bound of the number of slack variables, and $\bm{\eta}$ is the penalty parameters which are defined according to \cite{kochenberger2014unconstrained}. Then, the master problem can be represented in the QUBO formulation as
\begin{align} \label{qubo_min}
   \textbf{MP}_2:  \min_{\mathbf{v, w}} \quad
    & \Bar{\theta}(\mathbf{v}) +  f_Q^{\eqref{MP_1: original}}(\mathbf{w}) + f_Q^{\eqref{MP_1: feasiable}}(\mathbf{w})+ f_Q^{\eqref{MP_1: feasiable}}(\mathbf{w,v}). 
\end{align}

\subsubsection{Overall Algorithm}
\begin{algorithm}[t!]
\caption{The Proposed HQCBD Algorithm} 
\label{alg:1}
\begin{algorithmic}[1]
\REQUIRE Iteration index $\zeta = 1$, maximum iteration number $\zeta^{\text{max}}$, iteration threshold $\epsilon$, $UB^{(0)} = +\infty$, and $LB^{(0)}=-\infty$. Initialize $\mathbf{w}^{(0)}$.
\WHILE{$|\frac{\text{UB}^{(\zeta-1)}-\text{LB}^{(\zeta-1)}}{\text{UB}^{(\zeta-1)}}| > \epsilon$ or $\zeta < \zeta^{\text{max}}$}
        \STATE  $C^{\text{F}, (\zeta)}, C^{\text{O}, (\zeta)}, \text{UB}^{(\zeta-1)}$ $\leftarrow \textbf{SUB($\mathbf{w}^{(\zeta-1)}, \text{UB}^{(\zeta-1)}$)}$. 
        \STATE Add $C^{\text{F}, (\zeta)}$ or $C^{\text{O}, (\zeta)}$ to the master problem $\textbf{MP}_1$ and update $\text{UB}^{(\zeta)} \leftarrow \text{UB}^{(\zeta-1)}$. \label{alg1:add_cut}
        \STATE Set appropriate penalty numbers or arrays $\bm{\eta}$. \label{alg1:add_penalty}
        \STATE Reformulate both objective and constraints in the master problem $\textbf{MP}_1$ and construct the QUBO master problem $\textbf{MP}_2$ by using corresponding rules, i.e., (38)-(40). \label{alg1:reformulate}
        \STATE Solve the the master problem $\textbf{MP}_2$ by the quantum computer. \label{alg1:solve_master}
        \STATE Obtain the optimal solution $\mathbf{w}^{(\zeta)}$ and $\theta^{(\zeta)}$.
        \STATE  Update $\text{LB}^{(\zeta)} \leftarrow \theta^{(\zeta)}$. \label{alg1:update_lb}
    \STATE Set $\zeta \leftarrow \zeta$ + 1.
\ENDWHILE 
\item[]
\item[]

 \hspace{-1.5em} \nonumber \textbf{SUB($\mathbf{w}^{(\zeta-1)}, \text{UB}^{(\zeta-1)})$}:
     \STATE Fix $\mathbf{w}$ as $\mathbf{w}^{(\zeta-1)}$, and solve the subproblem $\textbf{SP}_2$ using a LP solver in the classical computer. \label{alg1:sub}
     \IF{subproblem $\textbf{SP}_2$ is infeasible}
      \STATE Obtain the feasibility cut $C^{\text{F}, (\zeta)}$. \label{alg1:feasible_cut}
      \ELSE
      \STATE Obtain the optimal solution $\mathbf{u}^{(\zeta)}$, optimality cut $C^{\text{O}, (\zeta)}$, and objective value $f^{\text{SP},(\zeta)}(\mathbf{w})$. \label{alg1:optimal_cut}
    \STATE $\text{UB}^{(\zeta-1)} \leftarrow \min\{\text{UB}^{(\zeta-1)}, f^{\text{SP},(\zeta)}(\mathbf{w})\}$.
     \ENDIF
\RETURN  $C^{\text{F}, (\zeta)}$, $C^{\text{O}, (\zeta)}$, and $\text{UB}^{(\zeta-1)}$.
\item[]
\ENSURE Optimal $\mathbf{w}^*, \mathbf{u}^*$. 
\end{algorithmic}
\end{algorithm}

The overall HQCBD algorithm is summarized in Algorithm \ref{alg:1}. The algorithm contains an iterative procedure.  We initialize the binary variables $\mathbf{w}$ and other parameters at the beginning. In the $\zeta$-th iteration, we first execute the SUB module. Specifically, we fix the binary variables $\mathbf{w}^{(\zeta)}$ and solve the subproblem $\textbf{SP}_2$ using the classical computer (Line \ref{alg1:sub}). If the subproblem $\textbf{SP}_2$ is infeasible, we can obtain a new feasibility cut  $C^{\text{F}, (\zeta)}$ (Line \ref{alg1:feasible_cut}). Otherwise, we obtain a new optimality cut $C^{\text{O}, (\zeta)}$ and update the performance upper bound $\text{UB}^{(\zeta-1)}$ by the optimal solution of the subproblem $\textbf{SP}_2$ (Line \ref{alg1:optimal_cut}). Then, we add the obtained feasibility cut $C^{\text{F}, (\zeta)}$ or optimality cut $C^{\text{O}, (\zeta)}$ to the master problem $\textbf{MP}_1$ (Line \ref{alg1:add_cut}). Next, we set the appropriate penalties and reformulate the master problem $\textbf{MP}_1$ into the QUBO formulation $\textbf{MP}_2$ (Lines \ref{alg1:add_penalty}-\ref{alg1:reformulate}). Finally, we utilize the quantum computer to solve the QUBO master problem $\textbf{MP}_2$ and update the performance lower bound $\text{LB}^{(\zeta)}$ (Lines \ref{alg1:solve_master}-\ref{alg1:update_lb}). This iteration procedure stops until the approximation gap $|\frac{\text{UB}^{(\zeta)}-\text{LB}^{(\zeta)}}{\text{UB}^{(\zeta)}}|$ is within a preset threshold $\epsilon$ or the maximal iteration index $L^{\text{max}}$ is reached. 

Note that the overall complexity analysis of our approach aligns well with the BD convergence analysis \cite{rahmaniani2017benders}. However, it is well known that the algorithmic complexity of QA is still an open question\cite{cubitt2015undecidability}. In the following Section~\ref{sec: experiment}, we conduct empirical experiments to demonstrate the superior performance of the HQCBD over the classical BD approach.

\subsection{Multi-cut Strategy}
\label{MC}
\begin{figure}[!t]
\centering
\subfloat[]{\includegraphics[scale=0.3]{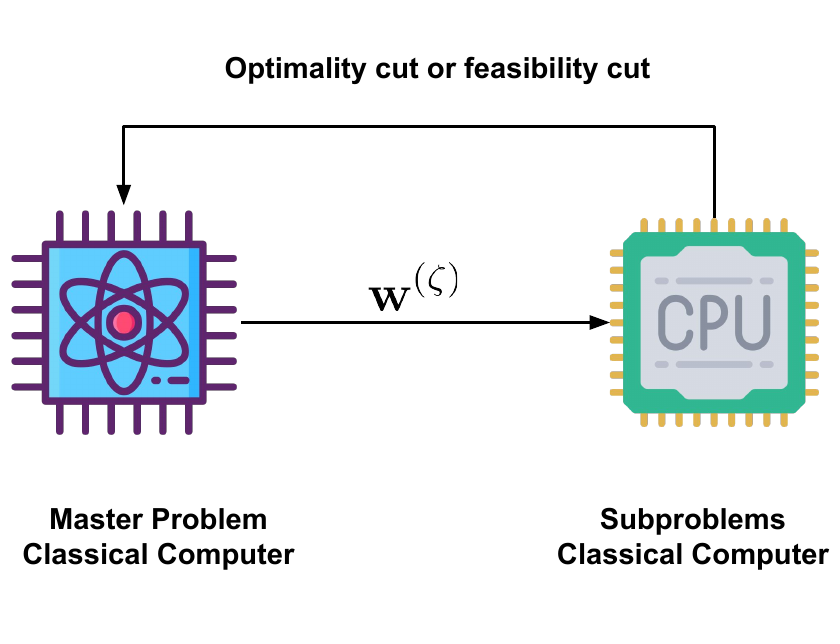}
\label{fig:cbd}}
\hfil
\subfloat[]{\includegraphics[scale=0.3]{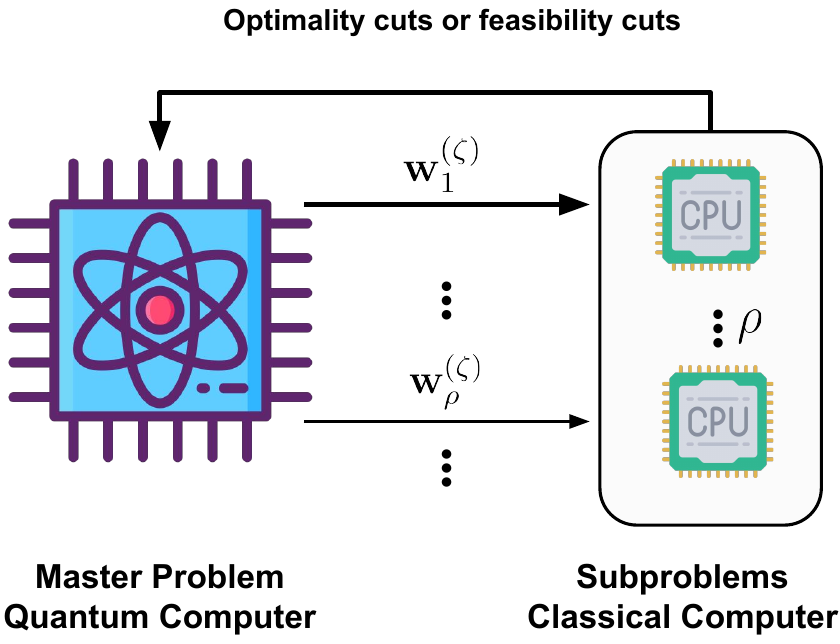}
\label{fig:MQBC}}
\caption{An overview of (a) Single-Cut HQCBD and (b) Multi-Cut HQCBD.}
\label{fig:cbd_mqbc}
\end{figure}

\begin{algorithm}[t!]
\caption{The Proposed Multi-Cut HQCBD Algorithm} 
\label{alg:2}
\begin{algorithmic}[1]
\REQUIRE Iteration index $\zeta = 1$, maximum iteration number $\zeta^{\text{max}}$, iteration threshold $\epsilon$, $UB^{(0)} = +\infty$, and $LB^{(0)}=-\infty$. Initialize $\rho$ feasible values as $\mathcal{X}^{(0)}=\{\mathbf{w}^{(0)}_i\}_{i=1}^{\rho}$.
\WHILE{$|\frac{\text{UB}^{(l-1)}-\text{LB}^{(l-1)}}{\text{UB}^{(l-1)}}| > \epsilon$ or $l < L^{\text{max}}$}
    \FOR{$\mathbf{w}_i^{(\zeta-1)} \in \mathcal{X}^{(\zeta-1)}$}
        \STATE  $C_i^{\text{F}, (\zeta)}, C_i^{\text{O}, (\zeta)}, \text{UB}^{(\zeta-1)}$ $\leftarrow \textbf{SUB($\mathbf{w}_i^{(\zeta-1)}, \text{UB}^{(\zeta-1)}$)}$.
    \ENDFOR
  \STATE Add all generated cuts to the master problem $\textbf{MP}_1$ and update $\text{UB}^{(\zeta)} \leftarrow \text{UB}^{(\zeta-1)}$.
        \STATE Set appropriate penalty numbers or arrays $\bm{\eta}$.
        \STATE Reformulate both objective and constraints in the master problem $\textbf{MP}_1$ and construct the QUBO master problem $\textbf{MP}_2$ by using corresponding rules, i.e., (38)-(40).
        \STATE Solve the master problem $\textbf{MP}_2$ by the quantum computer.
        \STATE Obtain $\rho$ feasible solutions ${\mathcal{X}}^{(\zeta)} \leftarrow \{\mathbf{w}^{(\zeta)}_i\}_{i=1}^{\rho}$ and  update $\text{LB}^{(\zeta)} \leftarrow \min \{\theta^{(\zeta)}_i\}_{i=1}^{\rho}$.
    \STATE Set $\zeta \leftarrow \zeta + 1$.
\ENDWHILE 
\ENSURE Optimal $\mathbf{w}^*, \mathbf{u}^*$..
\end{algorithmic}
\end{algorithm}

The primary bottleneck in classical BD is the time consumed in solving the master problems, which accounts for over $90\%$ of the total optimization time \cite{magnanti1981accelerating}. As illustrated in Fig.~\subref*{fig:cbd}, single-cut HQCBD just yields one feasibility cut or optimality cut at each iteration. It will need a large number of iterations to converge if the quality of the generated cuts is low. To address this problem, we first observe that QA can generate multiple feasible solutions at each iteration. This capability arises because QA employs qubits, which can explore numerous combinations of quantum states simultaneously by leveraging the superposition principle \cite{bharti2022noisy}. Consequently, quantum computers have the potential to further accelerate convergence by constructing multiple cuts. Based on this, we design a specialized quantum multi-cut strategy, which is shown in Fig.~\subref*{fig:MQBC}. The details of the multi-cut HQCBD are outlined in Algorithm \ref{alg:2}. The major difference between Algorithms \ref{alg:1} and \ref{alg:2} is that a set of feasible binary variables $\mathcal{X}$ is generated by solving the master problem on the quantum computer. For each feasible solution, a corresponding subproblem is solved on the classical computer to generate a feasibility or optimality cut. Then, all generated cuts are added to the master problem for the quantum computer to solve. This iteration procedure stops until the upper bound and lower bound converge.


\section{Numerical Experiments} 
\label{sec: experiment}
In this section, we evaluate our proposed algorithms through numerical experiments.  Due to the high cost of QPU utilization and time limitations for the developer, our experiments are limited to a small-scale setting. However, even with these hardware limitations, our results clearly demonstrate the immense potential of this technology for the future. We implement both HQCBD and BD in Python 3.10. Specifically, classical MILP and LP problems are solved using Gurobi \cite{Gurobi}. These classical algorithms are conducted on a server equipped with a 4.2 GHz AMD Ryzen Threadripper PRO CPU and 512 GB of RAM. On the other hand, the master problem $\textbf{MP}_2$ is solved by the real-world D-wave Advantage quantum computer, which has over 5,000 qubits and 35,000 couplers based on the Pegasus topology \cite{dwave}.

\begin{table}[t] 
\caption{Simulation Parameters}
\label{talbe: simulation parameters}

\centering
\begin{adjustbox}{width=0.78\columnwidth,center}
\begin{tabular}{|c |c |}
\hline
Parameters & Values  \\
\hline
Transmission power of satellite for S2S link & 20\si{W} \\ 
\hline
Transmission power of satellite for S2U link & 20\si{W} \\
\hline
Transmission power of user for U2S link & 1\si{W} \\
\hline
Combined antenna gain\footnotemark of U2S/S2U link   & 42\si{dBi} \\
\hline
Combined antenna gain of S2S link & 52\si{dBi} \\
\hline
Carrier center frequency of U2S/S2U  link & 30\si{GHz} \\
\hline
Carrier center frequency of S2S  link & 2.2\si{GHz} \\
\hline
\end{tabular}
\end{adjustbox}

\end{table}
\footnotetext{Combined antenna gain is the product of the transmit and receive antenna gains.}

\subsection{Simulation Setup}
\label{subsec:simulation step}
We consider an NFV-enabled SIN scenario comprising $12$ satellites and $8$ users. Specifically, $12$ satellites are arbitrarily selected from the Iridium Next constellation and are distributed over three polar orbits at a height of 781 \si{km}. $4$ users are located in the source city, San Antonio, USA $(\ang{29.42} N, \ang{98.49} W)$. Each user sends a computational task to its paired user in the destination city, Seattle, USA $(\ang{47.60} N, \ang{122.33} W)$, via the SIN. In other words, the number of task flows $L=4$. To simulate this time-varying SIN, we leverage the Satellite Communication Toolbox \cite{MATLAB_Satellite} and the two-line element (TLE) data \cite{CELESTRAK}. 

In this NFV-enabled SIN scenario, we assume that there are four service functions (i.e., $\mathcal{F}=\{f_1,\cdots, f_4\}$). Each task flow consists of 2 required service functions as SFC, which are randomly chosen from the total function set $\mathcal{F}$. The time horizon is set as $T = 300\si{s}$ and the time slot duration is $\delta = 10\si{s}$.  Subsequently, 6 satellites are randomly selected as function nodes, each capable of being deployed with two distinct functions. Besides, according to \cite{jia2020joint, jia2021toward}, we set the channel parameters as $L_l = -23 \si{dB}$, $k_B = 1.38 \times 10^{-23} \si{J/K}$, $T_s = 1000\si{K}$, $M=5\si{dB}$. The bandwidth $B_{i,j}$ for both S2U and U2S links is set to 30 \si{MHz}. The computation scaling factor $\beta_{\xi_{f_{l(k-1)}}, \xi_{f_{l(k)}}}$ is set within the range  $ [0.8, 1.2]$. Each satellite is configured with a storage capacity $C(v_i^t, v_i^{t+1}) = 50\si{Mb}$, and a computation capacity $C_i^t=1500 \si{Mbit/s}$.

For the setup of quantum computing, the continuous variable $\theta$ in master problem $\textbf{MP}_1$ is discretized by 20 binary variables. The problem was embedded into the physical QPU graph using the minor embedding by default settings \cite{dwave}. For all problems submitted to the QPU, the anneal-read cycle was repeated 1000 times. The rest of our simulation parameters, unless otherwise stated, are given in Table~\ref{talbe: simulation parameters}.

 Our evaluation comprises two parts. First, we assess the performance of the proposed single and multi-cut HQCBD algorithms by comparing them with the well-known classical approach, branch-and-bound (BnB) \cite{lawler1966branch, boyd2007branch, morrison2016branch}, which solves the master problem within the BD framework. 

Second, we evaluate the performance of the proposed U-VNF-R scheme against the following three baselines:
\begin{enumerate*}
 \item{Limited VNF deployment scheme (U-LVNF-R) :} In this approach, each function node provides only one function. This approach is the traditional approach, which is widely used in the VNF area \cite{jia2018joint, jia2020virtual,xu2023delay}.
  \item{Fixed SFC service scheme (U-FVNF-R):} This approach is very similar to that proposed in the previous works \cite{yang2021maximum}. Each function within a mission flow is served by a specific fixed function node.
  \item{Heuristic user association scheme (HU-VNF-R):} Each user is connected to the satellite with the strongest signal (i.e., maximum signal-to-noise ratio).
\end{enumerate*} Note that the baselines also leverage the proposed algorithm to optimize other variables.

\begin{figure}
\centering
  \centering
  \includegraphics[scale=0.4]{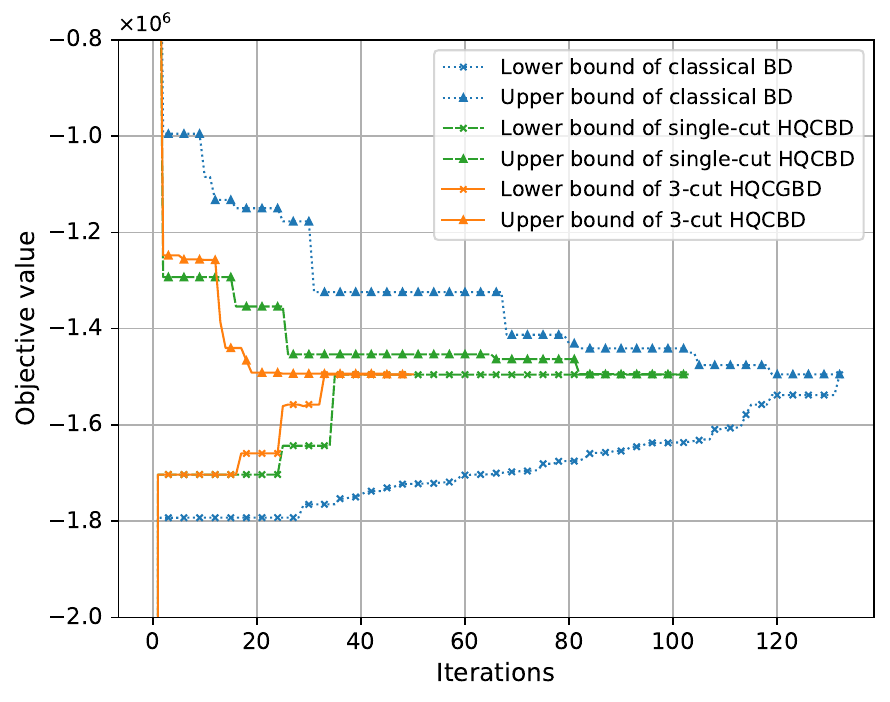}
  \caption{Convergence performance of different HQCBD strategies compared to the classical BD approach for solving problem  $\textbf{P}_1$ ($L=4, B_{i,j}=30\si{MHz}, C(v_i^t, v_i^{t+1})=50\si{Mbit}, C_i^t=1500\si{Mbit/s}$).}
  \label{fig:convergence}
\end{figure}

\begin{table}[t]
\centering

\caption{Solver accessing time of BD and multi-cut HQCBD strategy.}
\label{tab:compare}
\begin{tabular}{|>{\centering}p{3cm}|p{1.5cm}|p{1.5cm}|p{1cm}|}
\hline
\multirow{3}{*}{Algorithm}  & \multicolumn{3}{c|}{Solver Accessing Time (\si{ms})}                                                                            \\ \cline{2-4} 
                & \multicolumn{1}{>{\centering}p{1.5cm}|}{Max./Min.} & \multicolumn{1}{>{\centering}p{1.5cm}|}{Mean./Std.}  &  \multicolumn{1}{>{\centering}p{1cm}|}{Total}\\ 
                  \hline
   Classical BD      & \multicolumn{1}{c|}{91.70/0.31}  & \multicolumn{1}{c|}{35.03/24.75} &  \multicolumn{1}{c|}{4624.03}\\ \hline
   Single-cut HQCBD      & \multicolumn{1}{c|}{8.75/4.36}  & \multicolumn{1}{c|}{8.20/1.43} &  \multicolumn{1}{c|}{904.81} \\ \hline
   3-cut HQCBD      & \multicolumn{1}{c|}{8.74/4.37}  & \multicolumn{1}{c|}{8.13/1.50} &  \multicolumn{1}{c|}{414.93} \\ \hline
\end{tabular}
\end{table}

\subsection{Comparison of Proposed HQCBD and BD}
\label{subsec: Comparison of Proposed HQCBD and BD}
In this part, we study the convergence properties of the proposed HQCBD. From Fig.~\ref*{fig:convergence}, we observe that both the classical BD and HQCBD approaches can converge at similar iterations. This result verifies that our proposed HQCBD algorithm is mathematically consistent with the classical BD algorithm. In other words, if the classical BD algorithm can solve a problem, our proposed HQCBD algorithm can at least achieve the same result. Moreover, the 3-cut HQCBD converges much faster compared with single-cut HQCBD and classical BD. The reason is that we can improve the obtained lower bounds through the multi-cut strategy. Specifically, the 3-cut HQCBD can reduce the number of required iterations by $61.36\%$ and $50.96\%$ compared with the classical BD and single-cut HQCBD, respectively.

\begin{figure}[t]
     \centering
     \includegraphics[scale=0.4]{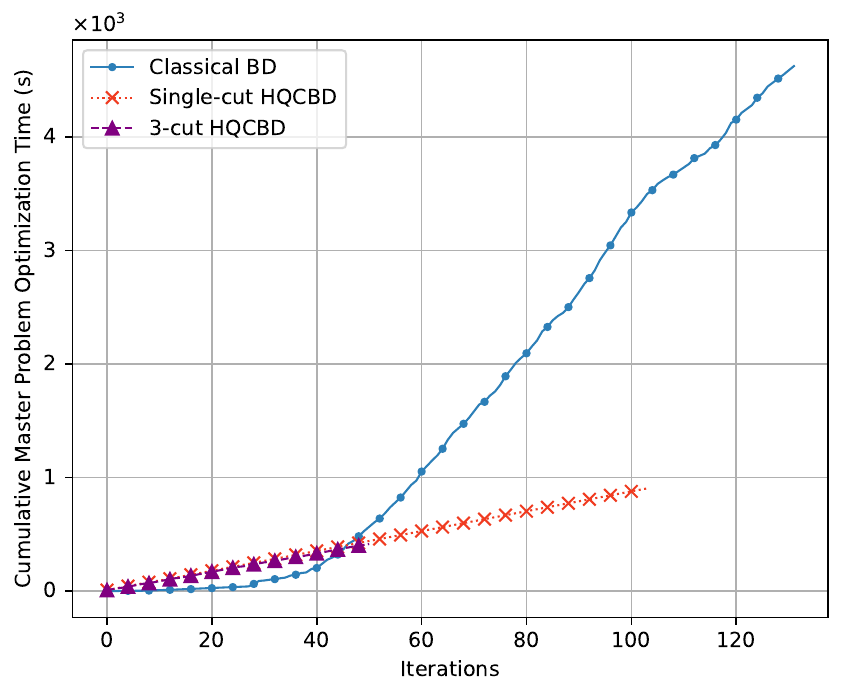}
      \caption[Caption for LOF]
      {The cumulative solver accessing time of master problems for the classical BD and different multi-cut HQCBD strategies\footnotemark. ($L=4, B_{i,j}=30\si{MHz}, C(v_i^t, v_i^{t+1})=50\si{Mbit}, C_i^t=1500\si{Mbit/s}$).}
      \label{fig:time}
\end{figure}

Additionally, we also investigate the running time of the algorithms. Note that although each iteration of 3-cut HQCBD requires processing multiple subproblems, the complexity of each subproblem is the same as that of the subproblem in the classical BD, and they can be executed in parallel. Since we only have access to the remote quantum computer, we ensure a fair comparison by evaluating the performance of the algorithms solely based on the actual solver access time for the master problems\footnote{The solver accessing time is the time required by the QPU solver or local solver, excluding other overheads such as variable setup latency, network transmission latency, etc.}. As demonstrated in Fig.~\ref{fig:time}, the single-cut HQCBD master problem's cumulative solver accessing time increases linearly w.r.t. the iteration number, while the classical BD master problems' cumulative solver accessing time increases quadratically w.r.t. the iteration number. Before the $47$-th iteration, the classical BD outperforms the single-cut HQCBD. However, as we keep adding Benders' cuts to the master problem in each iteration, the master problem becomes more and more complex. The quantum computer shows superiority in solving these large-scale MILP problems. In particular, the proposed single-cut and 3-cut HQCBD can save up to $80.43 \%$ and $91.02 \%$ solver accessing time of the master problem compared with the classical BD, respectively.  

\footnotetext{The solver accessing time of classical BD is obtained by averaging over 10 realizations.}

Finally, we show the performance details in Table~\ref{tab:compare}. We can observe that the quantum algorithms (i.e., single-cut and 3-cut HQCGB) have a stable computation performance as they have a much smaller standard deviation of the master problem's solver accessing time than the classical BD. In summary, the proposed multi-cut HQCBD outperforms classical BD in terms of both convergence speed of iterations and cumulative master problems' solver accessing time.


\subsection{Impact of Parameters}
\label{sec:Impact_of_parameters}
\begin{figure}[t]
     \centering
     \includegraphics[scale=0.4]{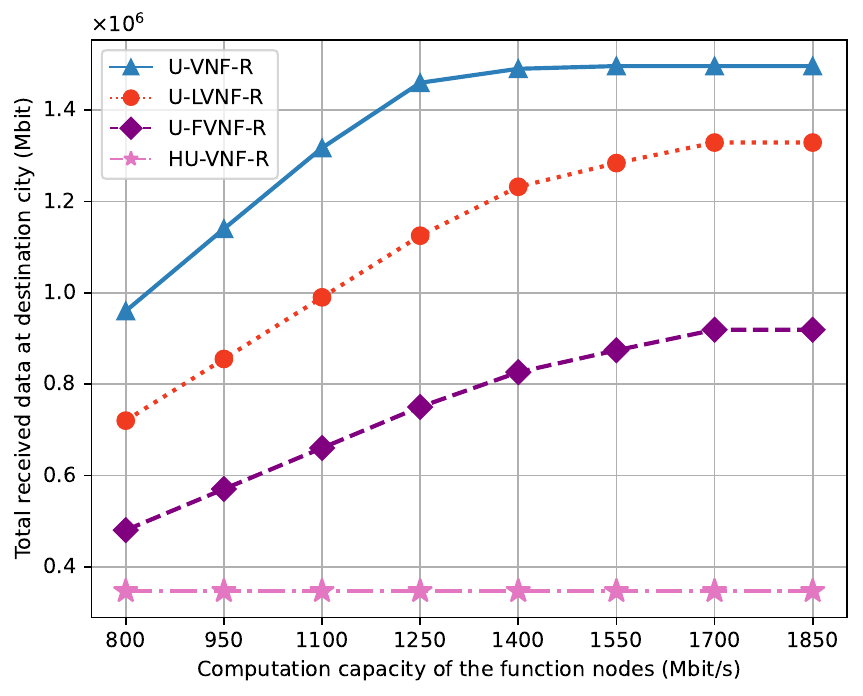}
      \caption{The impact of function node's computation capacity on network performance ($L=4, B_{i,j}=30\si{MHz}, C_i^t=50\si{Mbit}$).}
      \label{fig:computation}
\end{figure}
\textbf{\emph{Impact of function node's computation capacity $\bm{C_i^t}$.}} In this part, we investigate how the total received data changes as the function nodes' computation capacity $C_i^t$ increases from $800$ to $1850$ \si{Mbit/s}. As shown in Fig.~\ref{fig:computation}, the total received data increases as the function nodes' computation capacity increases. The reason is that the computation capacity of function nodes is one of the bottlenecks. With a larger computation capacity of function nodes for a given network size, more data can be processed within this network. Moreover, it can be seen that our proposed U-VNF-R scheme significantly outperforms the other three baselines, which demonstrates the effectiveness of our joint optimization design. Besides, we can also observe that the HU-VNF-R scheme is the worst compared with the other two schemes. The reason is that even though ground users connect with the satellite providing the strongest signal, the task flows are constrained by the computation and communication capacity of the associated satellite, which significantly degrades network performance. This further confirms the importance of optimizing user associations in this network.

\begin{figure}[t]
     \centering
     \includegraphics[scale=0.4]{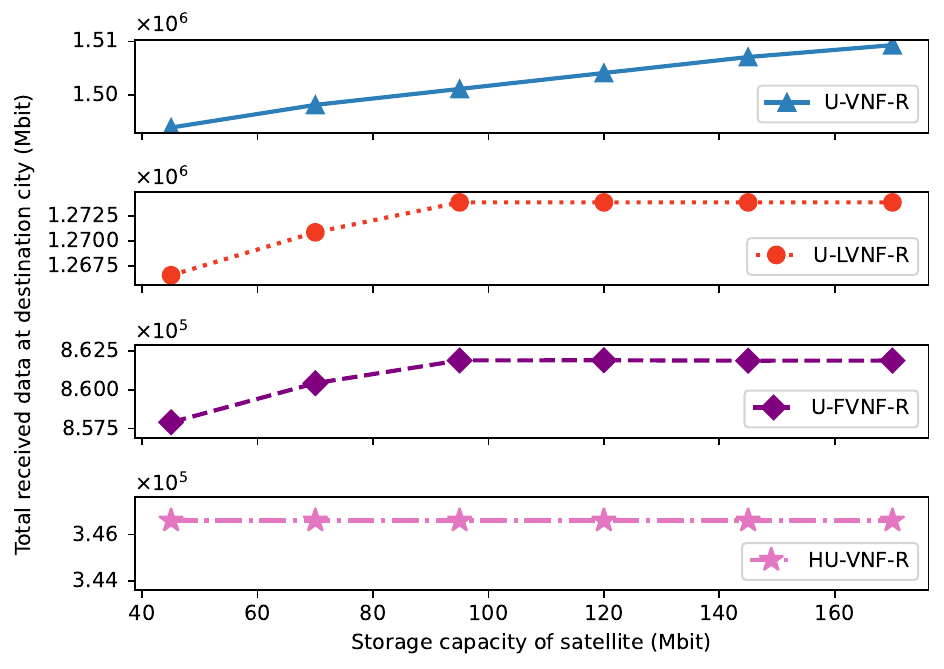}
      \caption{The impact of satellite's storage capacity on network performance ($L=4, B_{i,j}=30\si{MHz}, C_i^t=1500\si{Mbit/s}$).}
      \label{fig:storage}
\end{figure}

\begin{figure}[t]
     \centering
     \includegraphics[scale=0.4]{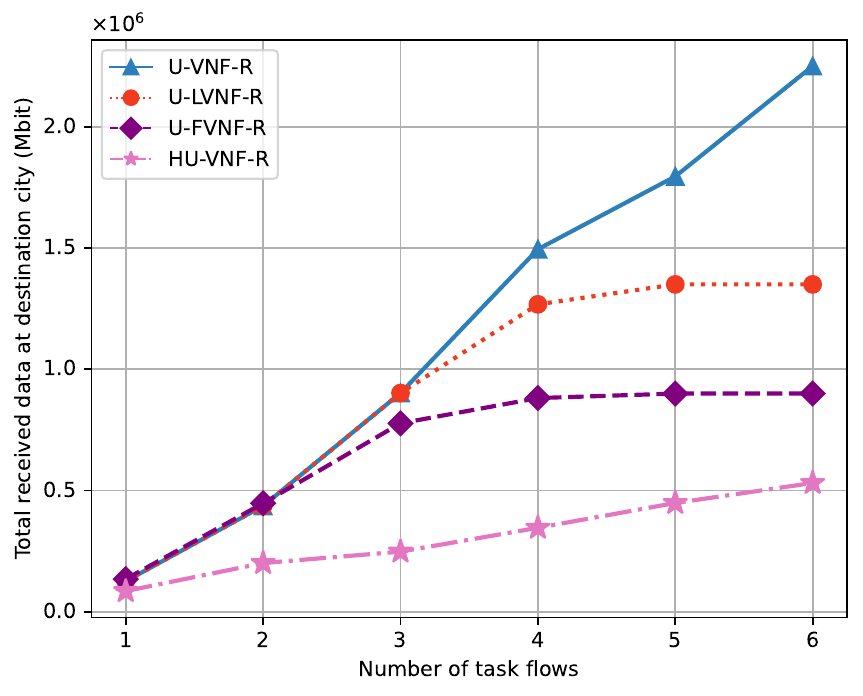}
      \caption{The impact of the number of task flows on network performance ($ B_{i,j}=30\si{MHz}, C(v_i^t, v_i^{t+1})=50\si{Mbit}, C_i^t=1500\si{Mbit/s}$).}
      \label{fig:task_flow}
\end{figure}

\begin{figure}[t]
     \centering
     \includegraphics[scale=0.4]{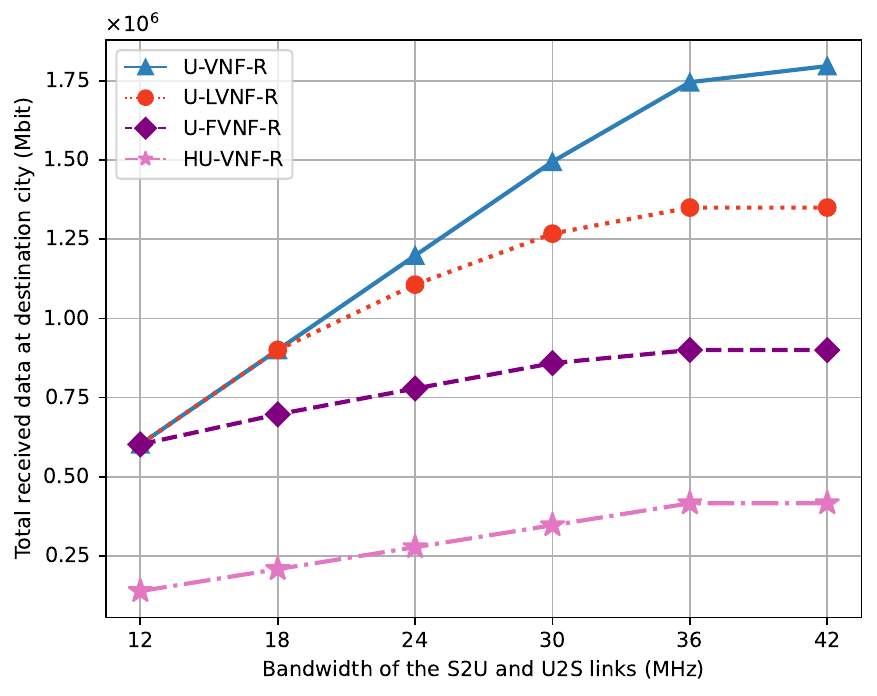}
      \caption{The impact of the bandwidth of S2U and U2S links on network performance ($L=4, C(v_i^t, v_i^{t+1})=50\si{Mbit}, C_i^t=1500\si{Mbit/s}$).}
      \label{fig:bandwidth}
\end{figure}


\textbf{\emph{Impact of satellite's storage capacity $\bm{C(v_i^t, v_i^{t+1})}$.}} In this part, we compare the total received data of our proposed U-VNF-R scheme with the baseline schemes regarding the storage capacity of satellites. As illustrated in Fig.~\ref{fig:storage}, our U-VNF-R scheme achieves the highest total received data compared to the baselines across various sizes of satellite storage capacity. Moreover, the total received data of all schemes increases as the satellite storage capacity increases, except that the total received data of the HU-VNF-R scheme remains stable. Essentially, enhanced data storage capacity allows for holding more data until it is transmitted to a suitable next node. Compared to U-LVNF-R and U-FVNF-R, a larger storage capacity significantly enhances network performance for the proposed U-VNF-R. This improvement is due to U-VNF-R's greater flexibility in selecting the execution locations for the required function servers of the task flows, thereby fully utilizing the storage capacity of each node.

\textbf{\emph{Impact of the number of task flows $\bm{L}$.}} Next, we study how the number of task flows affects the total received data. To that end, we increase the number of task flows from 1 to 6. Each task flow comprises 2 required service functions, forming an SFC, which are randomly selected from the total function set $\mathcal{F}$. From Fig.~\ref{fig:task_flow}, we observe that the total received data for the U-LVNF-R and U-FVNF-R schemes initially increases and then stabilizes as the number of task flows increases. This is because these schemes cannot fully utilize the computation and communication capacities of each satellite. Consequently, the network performances of these two schemes are inferior to that of U-VNF-R.

\textbf{\emph{The impact of the bandwidth of S2U and U2S links $\bm{B_{i,j}}$.}} Then, we investigate how the total received data changes as the bandwidth of S2U and U2S links $\bm{B_{i,j}}$ increases from $12$ to $42$ \si{MHz}. As shown in Fig.~\ref{fig:bandwidth}, the total received data increases as the bandwidth of S2U and U2S links increases. The increasing bandwidth of S2U and U2S links results in a larger streaming task flow into the SIN, allowing it to process and deliver more data to the destination users. Notably, the performance of U-VNF-R is significantly higher than that of the other three baselines. This superiority is due to U-LVNF-R and U-FVNF-R being constrained by their inflexibility in serving the functions of the task flow, while HU-VNF-R is limited by the computation and communication capacities of the associated satellite.

\section{Conclusion} 
\label{sec: conclusion}

In this paper, we have proposed a novel NFV-enabled SIN in which satellites cooperatively provide end-to-end communication and computation services for ground users. Based on MF-TEG, We have formulated a MILP that jointly optimizes the user association, VNF deployment, and the flow routing strategy, with the goal of maximizing the total processed data received by ground users. Then, we developed a hybrid quantum-classical solution method called HQCBD to solve it. To accelerate the HQCBD convergence, we have also designed a specialized quantum multi-cut strategy. The simulation results have demonstrated the advantages of our proposed multi-cut HQCBD in terms of iteration number until convergence and solver accessing time while ensuring optimality. This work is our first attempt to leverage quantum computing techniques for optimizing the user association, VNF deployment, and the flow routing strategy in the NFV-enabled SIN system. Since the proposed algorithm can efficiently address large-scale MILPs, it holds promise for various SIN applications, e.g., cache placement optimization problems. With the rapid development of quantum computers and increasing qubits \cite{dwave}, we believe that quantum-assisted optimization will play a significant role in the SIN field.

\bibliographystyle{IEEEtran}
\bibliography{IEEEabrv,ref}

\end{document}